\documentclass{article}
\usepackage[left=1in, right=1in, top=1in, bottom=1in]{geometry}
\usepackage[utf8]{inputenc}
\usepackage{subfig}

\usepackage{amsthm}
\usepackage{amssymb}
\usepackage{amsmath}
\usepackage{mathtools}
\usepackage{stmaryrd}
\let\olddotlessi\i
\usepackage[sanserif,full]{complexity}
\let\i\olddotlessi
\usepackage{csquotes}
\usepackage{enumitem}   
\usepackage{tikz}
\usepackage{graphicx}
\usepackage{xspace}
\usetikzlibrary{arrows.meta,matrix}
\usepackage[colorlinks=true,allcolors=blue,allbordercolors=white]{hyperref} 
\usepackage{caption}  
\usepackage{dirtytalk}
\usepackage{thm-restate}

\setlist[enumerate]{itemsep=\smallskipamount,parsep=0pt,label={\rm \roman*)}}
\setlist[itemize]{itemsep=\smallskipamount,parsep=0pt}

\usepackage[skip=10pt plus2pt]{parskip}
\makeatletter
\def\thm@space@setup{%
  \thm@preskip=\parskip \thm@postskip=0pt
}
\makeatother
\makeatletter
\renewenvironment{abstract}{%
  \if@twocolumn
    \section*{\abstractname}%
  \else
    \small
    \begin{center}%
      {\bfseries \abstractname\vspace{-.5em}\vspace{\z@}}%
    \end{center}%
    \quotation
    \setlength{\parskip}{8pt plus2pt}
  \fi}
  {\if@twocolumn\else\endquotation\fi}
\makeatother

\newcommand{\eps}{\varepsilon}

\newcommand{\defn}[1]{{\textit{\textbf{\boldmath #1}}}\xspace}
\renewcommand{\paragraph}[1]{\vspace{0.09in}\noindent{\bf \boldmath #1.}}

\usepackage{bbm}

\newcommand{\interior}[1]{ {\kern0pt#1}^{\mathrm{o}} }

\newcommand{\set}[1]{\left\{ #1\right\}}

\newcommand{\Q}{\mathbb{Q}}

\newcommand{\Z}{\mathbb{Z}}
\newcommand{\F}{\mathbb{F}}
\newcommand{\N}{\mathbb{N}}

\newcommand{\calK}{\mathcal{K}}

\usepackage{wasysym}
\usepackage{algorithm}
\usepackage{algpseudocode}

\newcommand{\floor}[1]{\left\lfloor #1 \right\rfloor}
\newcommand{\ceil}[1]{\left\lceil #1 \right\rceil}
\newcommand{\paren}[1]{\left( #1 \right)}

\newcommand{\bin}{\{0,1\}}

\usepackage[capitalise,nameinlink,noabbrev]{cleveref}
\crefname{equation}{}{} 
\crefname{enumi}{Step}{} 

\theoremstyle{definition}
\newtheorem{theorem}{Theorem}[section]

\newtheorem{definition}[theorem]{Definition}

\newtheorem{proposition}[theorem]{Proposition}
\newtheorem{lemma}[theorem]{Lemma}
\newtheorem{claim}[theorem]{Claim}
\newtheorem{corollary}[theorem]{Corollary}

\usepackage{mdframed}
\usepackage{framed}

\newcommand{\softo}[1]{\widetilde{O}\left( #1 \right)}

\usepackage{cryptocode}
\newclass{\CL}{CL}

\def\draft{1}
\newcommand{\nnote}[1]{\ifnum\draft=1\textcolor{purple}{[\textbf{Nathan:} #1]}\fi}
\newcommand{\rnote}[1]{\ifnum\draft=1\textcolor{teal}{[\textbf{Ryan:} #1]}\fi}
\newcommand{\anote}[1]{\ifnum\draft=1\textcolor{orange}{[\textbf{Adelmo:} #1]}\fi}

\newcommand\blfootnote[1]{%
  \begingroup
  \let\thefootnote\relax
  \footnotetext{#1}%
  \endgroup
}
\usepackage[
    backend=biber,
    style=alphabetic,
    maxnames=10,
    maxalphanames=10, 
]{biblatex}
\def \AC {{\sf AC}}

\def \poly {{\text{poly}}}

\def \eps {{\varepsilon}}

\def \NL {{\sf NL}}

\def \CL {{\sf CL}}

\def \ZPP {{\sf ZPP}}
\def \TC {{\sf TC}}
\def \NC {{\sf NC}}

\def \eps {\varepsilon}
\def \poly {\text{poly}}

\newcommand{\ringk}{\mathcal{K}} 

\DeclareMathOperator{\gtype}{gate\_type} 
\DeclareMathOperator{\incoming}{incoming} 
\DeclareMathOperator{\indeg}{in\_degree} 
\newcommand{\abs}[1]        {\left| #1\right|}
\newcommand{\ts}{\textsuperscript}

\usepackage{stmaryrd}
\usepackage{comment}
\usepackage{listings}

\newcommand{\vecx}{\textbf{x}} 

\renewcommand{\leq}{\leqslant}
\renewcommand{\geq}{\geqslant}

\title{Computing faster with a full memory:\\ Catalytic algorithms break time-space lower bounds}
\author{Adelmo Morrison Orozco\\MIT\\ \href{mailto:adelmo@mit.edu}{adelmo@mit.edu} \and Nathan Sheffield\thanks{Work supported in part by an NSF GRFP fellowship.}\\MIT\\\ \href{mailto:shefna@mit.edu}{shefna@mit.edu}
\and Ryan Williams\thanks{Work supported in part by NSF CCF-2127597 and NSF CCF-2420092.}\\MIT\\\ \href{mailto:rrw@mit.edu}{rrw@mit.edu} }
\date{}

\begin{document}

\maketitle
\blfootnote{\\ \textbf{AI disclosure:} We used several models to search for references, and to proofread the paper. 
}
\begin{abstract}
    We show that adding ``catalytic space'' (which starts full and must have its original contents restored at the end of computation) provably speeds up time-and-space-bounded computation.
    In a natural output model, with $O(\log(n))$ free space and $\poly(n)$ catalytic space, we give algorithms for sorting and the discrete Fourier transform running in time $n^{1+o(1)}$, and an algorithm for matrix multiplication running in time $n^{\omega + o(1)}$.
    In the same output model, without additional catalytic space, sorting and DFT are known to require time-space product $\Omega(n^2)$, and matrix multiplication is known to require time-space product $\Omega(n^3)$ \cite{borodin1980time, beame91, Yesha1984, DBLP:journals/jcss/Abrahamson91}.
    This is the first unconditional demonstration that (uniform, non-relativized) catalytic logspace algorithms can outperform their (non-catalytic) counterparts.

    We give several further results relevant to fast catalytic algorithm design.
    We observe that many applications of sorting, FFT, and fast matrix multiplication (e.g., {\sc kSUM}, integer multiplication, and clique counting) can be performed catalytically, giving catalytic logspace algorithms matching (up to subpolynomial factors) the best known unbounded-space runtimes.
    Furthermore, we give a fast and simple catalytic algorithm for evaluating low-depth threshold circuits, improving the previously known runtime bound from $\softo{|C|4^df}$ to $\softo{|C|2^d}$, where $|C|$ is the total size (i.e. number of wires) of the circuit, $d$ is the depth, and $f$ is the maximum fan-in. 
    
\end{abstract}

\newpage
\tableofcontents
\section{Introduction}

\paragraph{Catalytic computing} In the model of catalytic computing, introduced by Buhrman, Cleve, Kouck\'{y}, Loff, and Speelman~\cite{bckls14}, 
an algorithm with a small amount of memory (e.g. $O(\log(n))$-bits, where the input is of length $n$) is augmented with a large amount of ``dirty'' memory (e.g. $\poly(n)$ bits).
The clean memory is initially empty, but the dirty memory begins full of arbitrary bits the algorithm cannot choose.
The algorithm may use both the clean and dirty memory however it sees fit to solve the problem at hand, but it is required to set every bit of the dirty memory back to its initial value when it terminates.
(A standard analogy for dirty memory is that one ``borrows'' a friend's enormous hard drive full of precious digital memories in order to carry out a computation, but one does not want to lose a friend in the process, so one is careful to restore every bit of the hard drive when done.) 

The initial impetus for defining this model was to prove that dirty space is not helpful for computation.
This would make intuitive sense: in order to restore every bit of the initial dirty memory contents at the end of computation, one must maintain all of the information throughout the computation, so one cannot ``free up'' any space in general.
However, remarkably, Buhrman, Cleve, Kouck\'{y}, Loff, and Speelman eventually found ways to make use of dirty space, solving problems that we don't know how to solve using clean space alone. 
\cite{bckls14} defined $\CL$ (``catalytic logspace'') to be the class of problems solvable with $\poly(n)$ bits of catalytic space and $O(\log n)$ clean space, and showed the containments $\TC^1 \subseteq \CL \subseteq \ZPP$. In particular, $\NL$ and $\BPL$ are both contained in $\CL$, showing that $\poly(n)$ dirty space is sufficient to simulate both randomness and nondeterminism in the logspace setting.

This model has generated a large amount of excitement and work among complexity theorists in recent years. We do not aim to give an exhaustive survey here, but some highlights are:
\begin{itemize}
\item new algorithms placing more problems in $\CL$ (such as~\cite{DBLP:journals/cc/Pyne25,DBLP:conf/focs/AgarwalaM25,DBLP:conf/mfcs/AlekseevFMSV25,DBLP:conf/innovations/AgarwalaAV26}),
\item new characterizations demonstrating $\CL$ is ``robust'' under natural extensions of the model (such as~\cite{DBLP:conf/stoc/CookLMP25,DBLP:conf/innovations/FolkertsmaMST25,DBLP:conf/focs/KouckyMPS25}), and
\item new developments in classical space complexity leveraging catalytic ideas~\cite{cookmertz20,cookmertz24,williams25}.
\end{itemize}

\paragraph{Catalytic computation as a means of speeding up low-space algorithms} For all its successes in complexity theory, so far catalytic computing has not caught on as a major topic of study for the rest of core theoretical computer science. One major reason is that almost all positive results in catalytic computing have been solely focused on minimizing space, while disregarding \emph{running time}. 
For most TCS researchers, runtime is the gold standard of algorithm performance; space usage is important as a secondary optimization target, but an enormous runtime precludes usefulness. Most known catalytic algorithms are of limited interest from this perspective, as they involve \cite{bckls14}'s $\TC^1$ simulation, which has a prohibitively large runtime overhead. 
    (For instance, using standard reductions, the resulting $\CL$ algorithm for directed $st$-connectivity --- the canonical $\NL$-complete problem --- would have a runtime of $\widetilde\Theta\paren{n^8}$.)

There has been some recent effort to correct this: Cook and Pyne initiated this direction by giving $\softo{n^3m}$-time and $\softo{nm}$-time deterministic and randomized $\CL$ algorithms, respectively, for directed $st$ connectivity \cite{DBLP:conf/innovations/CookP26}.
Subsequent works have shown that most of the space usage in some existing time-space tradeoff upper bounds can be made catalytic \cite{chmel2026frontier}.

In this paper, we give strong additional evidence that dirty space can be useful for designing fast algorithms.
We implement several essential tools in modern algorithm design (most centrally, sorting, fast matrix multiplication, and fast Fourier transform) catalytically with little loss in runtime, using only $O(\log n)$ clean space.
Near-linear time sorting and Fourier transforms are known to be \emph{unconditionally impossible} using only $n^{1-\Omega(1)}$ workspace in total, so our results demonstrate that dirty space can yield strictly faster algorithms than would be otherwise possible (similar results are known for matrix multiplication)~\cite{borodin1980time, beame91, Yesha1984, DBLP:journals/jcss/Abrahamson91}. (Our algorithm's output is provided by XORing into a chunk of dirty space, as opposed to the separate-output-tape model under which those lower bounds are usually stated. However, we also show that catalytic algorithms 
have a strict \emph{unconditional speedup} in output models where the lower bounds apply.)

\subsection{Our Results}

\paragraph{Fast catalytic algorithms} We begin by showing that threshold circuit evaluation, from which (under derandomization assumptions) most known containments in $\CL$ follow, can itself be done substantially faster: a single-bit-output circuit of size $w$ and depth $d$ can be evaluated in catalytic logspace in time $\softo{w 2^d}$.\footnote{As is standard, we use $\widetilde O(T(n))$ as shorthand for $T(n) \cdot \poly(\log T(n))$.} (The algorithm of \cite{bckls14} has runtime $\widetilde{\Theta}(w f 4^d)$, where $f$ is the circuit's maximum fan-in.) Our algorithm is substantially simpler than that of \cite{bckls14}, relying on a single combinatorial trick as opposed to their algebraic approach. 

This approach can also be applied to circuits with \emph{multi-bit} outputs. When the output is too large to store in the machine's clean space, we say a catalytic algorithm ``computes'' a function $g$ if, instead of terminating with the dirty space restored, it terminates with $g(x)$ bitwise-XORed into an initial segment of the dirty space. (Note that the dirty space could then be restored by running the algorithm a second time. This output model is particularly natural for designing and composing catalytic algorithms, and hence is our default. See the end of this section, and \Cref{sec:prelims}, for discussion of alternative models that instead use a separate output tape, which is necessary for comparing to non-catalytic algorithms.) 
Our circuit evaluation result in this model is as follows:

\begin{restatable}{theorem}{circeval}\label{thm:circeval}
There exists a catalytic algorithm using $O(\log(w) + d)$ clean workspace and $\softo{w 2^d}$ time, which takes in a $d$-layer, $w$-wire circuit of symmetric gates, and an input to the circuit, and computes the circuit's output.
\end{restatable}

We then show several consequences of fast circuit evaluation:

\begin{corollary}\label{cor:nc1}
    Any problem with size-$S(n)$ (uniform) $\NC^1$ circuits has a time-$S(n)^{1+o(1)}$ catalytic algorithm using $O(\log(n))$ clean space.
\end{corollary}

\begin{corollary}\label{cor:sorting}
    There exists a catalytic algorithm using $O(\log(n + b))$ clean space and $bn^{1 + o(1)}$ time which takes as input a length-$n$ list of $b$-bit numbers, and computes the sorted list.
\end{corollary}

\cref{cor:nc1,cor:sorting} follow from direct applications of \cref{thm:circeval}.
We show that one can evaluate structured low-degree arithmetic circuits quickly, yielding several further algorithmic consequences:

\begin{theorem}\label{thm:informalmm}
    There is a catalytic algorithm which takes in a pair of integer matrices $A, B \in [\poly(n)]^{n\times n}$ and computes their product $AB$, using $O(\log n)$ clean space and $n^{\omega+o(1)}$ time.\footnote{See further results about matrix multiplications over modular rings in \Cref{sec:lincircs}.}
\end{theorem}

\begin{theorem}[Informal] \label{thm:informaldft}
    Over any ``reasonable'' ring $\ringk$, there exists a catalytic algorithm which takes in a vector $x \in \ringk^n$ and computes the discrete Fourier transform, using $O(\log n)$ clean space and $n \cdot \poly(\log n)$ time.\footnote{See \Cref{sec:lincircs} for a formal statement.}
\end{theorem}

Using near-linear-time sorting, FFT, and fast matrix multiplication as building blocks, we observe that we can obtain fast catalytic algorithms for a host of downstream applications, for instance:

\begin{restatable}[{\sc kSUM}]{corollary}{ksum} \label{cor:ksum}
Let $A_1, A_2, \cdots, A_k$ be lists of $n^{o(1)}$-bit integers. Let $n=\max_i \set{\abs{A_i}}$.
There exists a catalytic algorithm which, using $n^{\ceil{k/2}+o(1)}$ time, $n^{\ceil{k/2}+o(1)}$ catalytic space, and $O(\log n)$ clean space, decides whether there exists a $k$-tuple $(a_1,\cdots,a_k) \in A_1 \times \cdots \times A_k$  such that $a_1+\cdots+a_k=0$ .
\end{restatable}

\begin{restatable}[Triangle counting]{corollary}{triangle} \label{thm:triangle}
Let $G$ be an $n$-vertex simple undirected graph, given as an adjacency matrix.
There exists a catalytic algorithm which, using $n^{\omega+o(1)}$ time, $n^{\omega +o(1)}$ catalytic space, and $O(\log n)$ clean space, counts the number of triangles in $G$.
\end{restatable}

We similarly observe that we can attain fast catalytic algorithms for integer multiplication (\Cref{cor:intmult}), polynomial multiplication (\Cref{cor:polymult}), frequency moments (\Cref{cor:freqmoments}), small universe \#{\sc 3SUM} (\Cref{cor:3sumsmall}), string matching with wildcards (\Cref{cor:stringmatching}), $k$-path detection (\Cref{cor:paths}), $k$-cycle detection (\Cref{cor:cycles}), and $3k$-clique counting (\Cref{cor:cliques}).

\paragraph{Relationship to lower bounds} None of the above runtimes are known or believed to be possible for (non-catalytic) algorithms with only $n^{o(1)}$ free space. So, these are cases where dirty space seems to be genuinely necessary for fast algorithms. 
Indeed, recall that we know \emph{unconditional} time-space lower bounds for several of these problems (sorting, DFT, integer multiplication, and matrix multiplication). 
Our catalytic algorithms achieve runtimes for these problems that are therefore \emph{provably} impossible for non-catalytic algorithms.

At this point, one might justifiably object that we are not making a fair comparison. 
These lower bounds are for multi-output functions, where the algorithm is given a separate output tape to which it has limited access (e.g., it must output all of the bits in a single left-to-right pass --- this is known as the ``transducer'' output model).
However, our catalytic algorithms are instead designed to give the output by XORing into the catalytic tape.
So it is unclear whether the provable\footnote{We stress that for any applications to decision problems (e.g. 3SUM or triangle detection), this output distinction is irrelevant, because we could simply copy the first bit of the dirty space into the free space. So, even without any careful model bookkeeping, our results do show an advantage from dirty space, conditional on believable assumptions about the hardness of those problems. The interesting question here is whether we can get something \emph{unconditional}.} advantage is due to using dirty space to aid computation, or simply due to the different output model.

In \cref{sec:lbs}, we therefore give some apples-to-apples comparisons. 
We define the \defn{$k$-output-pass} output model, where the algorithm makes $k$ left-to-right passes over an output tape, to which it may XOR bits (but may not read existing values). (Note that the case $k=1$ corresponds to the transducer model.)
We then observe that all of our catalytic algorithms can be implemented in the $2$-output-pass model, and that all known time-space lower bounds (for non-catalytic algorithms) hold in the $k$-output-pass model for any constant $k$.
Thus, in this model, having dirty space is provably necessary for achieving fast runtimes.

It is interesting that our catalytic algorithms require two passes over the output, as opposed to just a single one. We show that this is indeed necessary, by noting the following generalization of the Borodin-Cook method~\cite{borodin1980time,beame91} to catalytic space algorithms:

\begin{theorem}\label{thm:lbs}
    For every \emph{catalytic} $1$-output-pass algorithm that takes in a length-$n$ list of $\log(n)$-bit integers and outputs the sorted list, we must have $ST \geq \Omega(n^2)$, where $S$ and $T$ are the clean space and runtime used by the algorithm, respectively.
\end{theorem}

It follows (from \Cref{thm:lbs} and our algorithmic results) that catalytic algorithms which make one pass over the output tape are strictly weaker than those which make two passes. However, we note that our results nonetheless show some separation between catalytic and non-catalytic algorithms even with only a single pass over the output (see \Cref{prop:transducerseparation}).

\subsection{Related Work}

As mentioned, few prior results in catalytic computing focus on time complexity, but two recent works have begun to explore this direction.

Cook and Pyne~\cite{DBLP:conf/innovations/CookP26} give elegant algorithms for catalytically solving directed connectivity (with $O(\log n)$ clean space) in randomized $\widetilde O(n\cdot m)$ time and deterministic $\widetilde O(n^3 \cdot m)$ time, and show how to estimate the probability that a $T$-step random walk ends at a given vertex in $\widetilde O(m T^2/\eps)$ time. These running times are close to the best-known running times achievable with large clean space, but are still at least one polynomial factor away.

Chmel, Dudeja, Kouck\'{y}, Mertz, and Rajgopal ~\cite{chmel2026frontier} studied how to transfer classical sublinear-space polynomial-time algorithms into the catalytic model. They give a catalytic algorithm for STCONN running in polynomial time, $O(\log n)$ clean space and $n/2^{\Theta(\sqrt{\log n})}$ catalytic space, matching the best-known space bound; with randomness, their algorithm runs in $O(m \cdot n^{2+\eps})$ time on $n$-node $m$-edge graphs. They also give polynomial time algorithms for edit distance, LCS, and discrete Fréchet distance using sublinear catalytic space, largely matching known non-catalytic space bounds. However, their focus is primarily on achieving the correct space frontier using only catalytic space, while retaining polynomial time bounds. In contrast, we focus on preserving running times to the greatest extent possible, replacing the usual space of clever classical algorithms with almost entirely catalytic space.

A related space-restricted model is the ``in-place'' model, where a read-write tape with input $x$ must be overwritten with the output $f(x)$, which has received recent attention in the computer algebra literature~(e.g. \cite{DBLP:journals/corr/Grenet25}) and in complexity theory~(\cite{in-place25}). The algorithms are usually allowed $O(\log n)$ bits of additional clean space, or sometimes $O(\log n)$ algebraic registers. In this model, among many other results, sorting can be computed in $O(n \log n)$ time~\cite{DBLP:journals/sigact/Pasanen99} and the Discrete Fourier Transform can be computed in $O(n \log n)$ time~\cite{in-place-FFT,DBLP:journals/siamsc/Temperton91}.
Another variant is the {\sc restore} model, wherein both the input and output tapes are read-write but the input tape must be restored to its initial contents at the end~\cite{DBLP:journals/talg/ChanMR18}. In this augmented variant, there exist additive accumulation algorithms over fields for the truncated Cooley-Tukey Fast Fourier Transform in $O(n \log n)$ time, Strassen's matrix multiplication algorithm in $O(n^{\log_2(7)})$ time \cite{DBLP:conf/issac/DumasG24}, efficient polynomial remainder~\cite{DBLP:conf/issac/DumasG24a}, and (more generally) linear accumulation of any in-place computable function \cite{DBLP:journals/jsc/DumasG26}.

We make reference to known time-space tradeoff lower bounds, and prove an extension to a catalytic model in \cref{sec:lbs}. 
These are among the small handful of unconditional lower bounds in complexity theory: 
a landmark result of Borodin and Cook shows that (in any ``reasonable'' model of computation, e.g. RAM Turing machines, with the transducer output model), the product of space overhead and runtime for any sorting algorithm must be at least $\Omega(n^2/\log(n))$ \cite{borodin1980time}, which Beame later improved to $\Omega(n^2)$ \cite{beame91}. Unconditional time-space-product lower bounds have since been shown for other problems, such as discrete Fourier transform (DFT), integer multiplication, and matrix multiplication~\cite{Yesha1984, DBLP:journals/jcss/Abrahamson91}.

\subsection{Proof Intuition}

We give here an overview of the proofs of the above results.
The main ideas are quite simple, although there are some technical details that one needs to be careful about.

\paragraph{Evaluating low-depth SYM circuits} The algorithm given in \cite{bckls14} for evaluating low-depth threshold circuits involved treating the dirty space as a sequence of ``register'' values over a finite field.
They show that, given an algorithm $A$ which results in a pair of values $a, b \in \F$ being added into two ``registers'' of the dirty space, one can use two calls to $A$ to get an algorithm which adds their product $ab$ into another register, while not storing anything except a global timestep counter in-between calls.
They give similar constructions for powering, and then use Fermat's little theorem to show that, given an algorithm $A$ which adds a vector of values $v \in \F^m$ to a sequence of registers, they can make four calls to $A$ and spend $\softo{wf}$ additional time in order to add the output of any single layer of threshold gates (with $w$ wires and maximum fan-in $f$) over $v$ to another sequence of registers. Recursing this algorithm then lets them evaluate low-degree circuits of threshold gates.

We use a similar recursive approach, but replace the algebraic methods with a single combinatorial trick, similar to one suggested by Cook \cite{cook2021borrow}. 
Our approach works for circuits of arbitrary symmetric gates (i.e., gates whose output only depends on the number of $1$s among input wires), including threshold gates.
The underlying idea is as follows.
Suppose you can toggle between seeing the value $\tau$ and the value $\tau + x$, for some $\tau, x \in \F_m$. 
Given a large ``helper'' array $H$ of dirty values, if you add $1$ to $H[\tau + x - i]$ for every $i \in S \subseteq \F_m$, then this will increment the value of $H[\tau]$ if and only if $x \in S$.
Using this fact, we can go from an algorithm adding a vector of bits, to an algorithm adding a $w$-wire layer of symmetric gates over those bits, using only two calls and $\softo{w}$ time overhead. Recursing this gives an evaluation algorithm running in time $\softo{w2^d}$ for circuits of depth $d$.

We then observe that this algorithm is near-linear time for circuits of depth $o(\log(n))$, so obtaining fast catalytic algorithms reduces to designing small, $o(\log(n))$-depth (uniform) circuits of symmetric gates.
In order to get such circuits, we take bounded fan-in circuit models ($\NC$ circuits, sorting networks) of depth $O(\log(n))$, and observe that we can collapse groups of a super-constant number of layers together into constant-depth SYM circuits at only a sub-polynomial blowup in size. This lets us directly convert known results in these models into fast catalytic algorithms.

\paragraph{Evaluating low-degree arithmetic circuits} The algebraic approach of \cite{bckls14} allows evaluating arithmetic circuits, with work which they note grows exponentially with the number of layers of the circuit. But, actually, the work needs only to grow exponentially in the number of layers of the circuit \emph{which involve non-linear product gates}: layers of the circuit which only compute linear functions (i.e. scalar product gates) needn't contribute.
The fast Fourier transform can be written as a small circuit of only linear gates, and any fast matrix multiplication algorithm can (without loss of generality) be written as a small circuit with a single layer of binary products and all other layers linear.
So, these algorithms can be simulated with minimal runtime overhead catalytically.

\paragraph{Time-space tradeoff lower bounds} The above algorithms are stated in a model where, instead of having a separate output tape, the catalytic algorithm is simply required to XOR the output into an initial portion of its catalytic tape.
We observe that any algorithm in this model can be directly simulated in the 2-output-pass model: in the first pass, output an initial portion of the catalytic tape; then, run the original algorithm until it gets to that specified intermediate point, and output the catalytic tape again in the second pass.
We therefore get fast algorithms for sorting, DFT, and matrix multiplication in this model.

On the other hand, we observe that any non-catalytic $k$-output-pass algorithm can be converted into a $1$-output-pass algorithm using only an $O(k)$-factor increase in space and an $\softo{1}$-factor increase in time, by holding $k$ different machine states in memory and simulating all $k$ passes in parallel. So, the lower bounds proven in the transducer (1-output-pass) model directly apply to the $2$-output-pass model as well, giving a separation between catalytic and non-catalytic algorithms.
In order to get a separation in the transducer model, we note that the above implies that the search problem ``given a list $L$, output $(L', L'')$ such that $L' \oplus L'' = \mathsf{SORTED}(L)$'' is possible with low product of time and clean space for catalytic transducers, but not for non-catalytic transducers.

Finally, to get a lower bound for the product of time and clean space against 1-output-pass catalytic algorithms, we directly adapt the Borodin-Cook branching program approach.
We consider \emph{catalytic} branching programs, which have a large number of start vertices, each with corresponding end vertices, and the requirement that computation from a given start vertex must always end at the corresponding end vertex.
As in existing lower bounds, we argue that, on a random input, most nodes of the branching program can't result in outputting too many values.
Then, we observe that \emph{every} path from a start to an end vertex must pass through many nodes that output many values. Since these paths have to be disjoint, this gives a large lower bound on the size of the branching program.

\section{Preliminaries}\label{sec:prelims}

We assume basic familiarity with computational complexity~\cite{arorabarak09} and arithmetic circuit complexity~\cite{DBLP:journals/fttcs/ShpilkaY10}.

\paragraph{Notation} We use $[N]$ to denote the set $\set{1,2,\cdots, N}$. We write $x \mod N$ to refer to the representative of $x$'s modular equivalence class in $[N]$, as opposed to the whole equivalence class. We use base 2 for the logarithm, $\log n:=\log_2(n)$. The set $\N=\set{1,2,\cdots}$ excludes 0.

\paragraph{Symmetric circuits} A function $f\colon \bin^m \to \bin$ is \emph{symmetric} if $f \circ \pi = f$ for every bitwise-permutation $\pi$. Equivalently, $f$'s output depends only on the number of $1$s in its input. (Examples of symmetric functions include $\texttt{AND}$, $\texttt{OR}$, or $\texttt{MAJORITY}$.) A \emph{symmetric circuit} is a circuit (of arbitrary fan-in) whose gates are all labeled with symmetric functions.

\paragraph{Catalytic algorithms} Consider an algorithm with read-only access to an input and read-write access to both a ``clean'' work-tape, as well as a ``dirty'' work-tape consisting of $d(n)$ bits. We say that such an algorithm is \emph{catalytic} if, on every input $x$ and every possible initial assignment $z \in \{0,1\}^{d(n)}$ to dirty space, the algorithm halts with dirty space set to $z$. We design our algorithms under the assumption that the model has RAM access to all of these tapes. 

\paragraph{A note on catalytic space complexity} It is known (see e.g. \cite{fortnow05}) that any RAM computation using time $T=T(n)$ accessing memory cells up to index, say, $2^{O(T)}$ can be converted to a time $T \cdot \poly(\log T)$ equivalent algorithm using $O(T \log T)$ contiguous bits of space. However, we do not yet know of a generic way to perform this conversion for \emph{catalytic} RAM algorithms running in time $T$. For this reason, we explicitly state the catalytic space bounds for the algorithms throughout this paper. All our algorithms have the feature that the catalytic space complexity is at most the runtime (up to polylog factors).

\paragraph{Low-space output models} For space-bounded computation of multi-bit functions, it is crucial to only allow restricted access to the output tape to retain a fair measure of the space. There are several possible ways to restrict the access, which we briefly survey, that could be of material consequence when measuring runtime.\footnote{All of the following notions are equivalent with respect to space, disregarding time complexity, by simply recording all the changes applied to the $i$\ts{th} bit of the output tape (for each $i$ at a time) and recomputing as necessary.}
For all the subsequent definitions, we say TM $M$ computes function family $f_n \colon \set{0,1}^n \to \set{0,1}^{\star}$ if, for all $x \in \set{0,1}^n$, whenever the input tape is initialized with $x$ the machine halts with $f(x)$ on the output tape.

The most restricted model is the bit-by-bit one-way output tape model, where the machine must \emph{commit} to the value of the $i$\ts{th} output bit before continuing to the $(i+1)$\ts{st} bit.
\begin{definition}[Transducer Output Tape] \label{def:transducer-output}
    A space $s=s(n)$ \textbf{transducer TM} $M$ computing a function family $f_n \colon \set{0,1}^{n} \to \set{0,1}^{\star}$ has access to:
    \begin{enumerate}
        \item An $n$-bit input tape with a read-only random-access head. 
        \item An $s$-bit work tape with a read-write random-access head.
        \item An output tape initialized to all zeros with a write-only one-way head, which starts in the leftmost cell and can only move right by one cell. Once the machine halts, the output is defined as all the bits up to the head's current position. (The tape is at most $2^{O(s)}$ bits long.)
    \end{enumerate}
\end{definition}
As the output tape is initialized to zero, moving right by one cell without overwriting the previous one is implicitly equivalent to writing a $0$ symbol to that previous cell, in the above definition. Observe both the input and work tapes are random access --- only the output tape is restricted to be one-way access. While the access is one-way, \Cref{def:transducer-output} does not restrict it to be write-once too (though, without loss of generality, this does not make a difference). Observe that this definition allows the output length to depend on the input itself, not only the input length.

All time-space lower bounds for multi-bit problems appealing to the branching program type defined in the Borodin-Cook argument \cite{borodin1980time} apply only to the ``append-only'' output form. This append-only notion, where we can choose to append some bit to the final output, is equivalent to \Cref{def:transducer-output}.
As the time-space lower bounds of \cite{Yesha1984,borodin1980time,DBLP:journals/jcss/Abrahamson91,beame91} are proven on $R$-way branching programs, a non-uniform analogue of the append-only transducer, they have only been shown to hold for the output tape model of \Cref{def:transducer-output}.\footnote{Note that a time-space lower bound on non-uniform $R$-way branching programs could be mildly stronger than a time-space lower bound for the transducer model. Aside from being non-uniform, an $R$-way branching program models input query access operations as unit cost and all other operations as \emph{free}. Thus, these time-space lower bounds apply for any space-bounded program with a ``reasonable'' measure of time. Furthermore, each input cell is assumed to lie in $\set{0,\cdots,R-1}$ instead of just the boolean set $\set{0,1}$.}

In contrast to clean space algorithms, catalytic ones suggest a different natural output model. Instead of placing the output into an append-only output tape, XOR the output into a chunk of your dirty space. This output model allows for \emph{composition} of catalytic algorithms which read their input by having it XORed into dirty memory. As these changes to the catalytic tape can be reverted by running the algorithm again, there is no danger of irreparably altering the catalytic tape.
\begin{definition}[Dirty-XOR-output model] \label{def:dirty-xor-output}
    A clean space $s=s(n)$, catalytic space $c=c(n)$ \textbf{dirty-XOR-output TM} $M$ computing a function family $f_n \colon \set{0,1}^{n} \to \set{0,1}^{\star}$ has access to:
    \begin{enumerate}
        \item An $n$-bit input tape with a read-only random-access head. 
        \item An $s$-bit clean work tape with a read-write random-access head.
        \item A $c$-bit catalytic work tape (arbitrarily initialized) with a read-write random-access head. Given input $x$, let $k=\abs{f(x)}$ and $\tau$ be the first $k$ bits of the dirty tape at initialization. At the end of the computation, the first $k$-bit dirty space chunk should be $\tau \oplus f(x)$.
        Each of the other remaining bits must be restored exactly.
    \end{enumerate}
\end{definition}

The above is restricted only to algorithms with access to dirty space. This motivates a definition of a useful output model applicable to both clean and catalytic space algorithms, in which we can make a comparison between them. The following output definition retains a similar sequential access restriction as \Cref{def:transducer-output} while incorporating the XOR-access feature of \Cref{def:dirty-xor-output}.

\begin{definition}[$k$-output-pass model] \label{def:kpass-xor-output}
    A space $s=s(n)$ \textbf{$k$-output-pass TM} $M$ computing function family $f_n \colon \set{0,1}^{n} \to \set{0,1}^{\star}$ has access to:
    \begin{enumerate}
        \item An $n$-bit input tape with a read-only random-access head. 
        \item An $s$-bit work tape with a read-write random-access head.
        \item An output tape, initialized to all zeros, with the following write pattern: $k$ passes of a one-way XOR head. One pass to the output tape consists of a one-way (XOR) head, starting at the leftmost cell and moving right by one cell, that is capable of XORing any value $b \in \set{0,1}$ to the current cell. To conclude one pass, the head position is reset to the leftmost cell --- this resetting is only done (at most) $k-1$ times. Once the machine halts, the output is defined as all the bits up to the \emph{rightmost} accessed head position over the entire computation. (The tape is at most $2^{O(s)}$ bits long.)
    \end{enumerate}
\end{definition}
Note this definition does not allow for overwriting bits of the output tape. As XORing $b=0$ into any bit keeps that bit the same and XORing by $b=1$ flips it, we can equivalently think of \Cref{def:kpass-xor-output} as allowing the machine to flip bits of the output tape in a multi-pass fashion. As bit-flips (XORs) can be undone by simply flipping (XORing) again, any machine in the above model will only modify the output tape \emph{reversibly} --- run the machine again to undo.

Observe that for $k=1$ the $k$-output-pass model (\Cref{def:kpass-xor-output}) is, in fact, equivalent in power to the transducer model (\Cref{def:transducer-output}), so the transducer is a special case of the $k$-pass model.\footnote{Indeed, on an output tape initialized to zeros, XORing bit $b \in \set{0,1}$ once onto the tape is equivalent to overwriting by $b$.}

\paragraph{Hash functions}
We define perfect hash families and note small such families can be constructed efficiently. These are used to obtain $k$-path and $k$-cycle algorithms in \Cref{sec:mm}.
\begin{definition}[Perfect Hashing] \label{perfect-hash-def}
    Fix a universe $U$. A $k$-perfect hash family $\mathcal{H}_{U,m}$ is a set of functions of the form $h \colon U \to [m]$ with the following property. For any set $S \subseteq U$ with $\abs{S}=k$ there exists an $h \in \mathcal{H}_{U,m}$ such that $h$ maps $S$ injectively onto $[m]$.
\end{definition}

\begin{lemma}[Uniform perfect hash, full table] \cite{MR1059653,MR819156} \label{perfect-hash}
    Let $n,k$ be positive integers. There exists a $k$-perfect hash family mapping from universe $[n]$ to $[m]$ with the following properties.
    \begin{itemize}
        \item Each function in the family is specified using $b=O(k + \log \log n)$ bits.
        \item We can iterate through all functions in the family in $2^{O(k)} \poly(\log n)$ time and $O(\log n + k)$ space.
        \item For input $x \in [n]$ and function $h$ in the family, we can compute $h(x)$ in $O(\log (n+k))$ space and $\poly(\log(n + k))$ time.
    \end{itemize}
\end{lemma}

\paragraph{Uniformity}
We define a natural notion of uniform arithmetic circuits and boolean circuits that will be useful for our catalytic algorithm applications. In short, a circuit family $\{C_n\}$ is uniform in our sense if the gate and wire information can be computed in polylog time and logspace, and (when arithmetic) every bit of every scalar on every wire can be determined in polylog time and logspace. \footnote{For circuits, this is distinct from standard logspace uniformity in that the usual definition imposes no restriction on the runtime (aside from the polynomial bound implied by any logspace algorithm). Our definition ensures we can simulate the circuits with little unnecessary time overhead.}

Let $\{C_n\}$ be a polynomial-size arithmetic circuit family, and fix a ring $\ringk$.
Let $S^{gates}(n)$, $S^{wires}_{\ell}(n)$, and $S^{wires}(n)$ be the total number of gates in $C_n$, the total number of wires in layer $\ell$ of $C_n$, and the total number of wires in $C_n$, respectively.
\begin{definition}[Succinctly-uniform arithmetic circuits] \label{arithmetic-uniform-circuit} 
We say that arithmetic family $\{C_n\}$ is {\bf succinctly-uniform} if there is a $\poly(\log n)$-time, $O(\log n)$-space algorithm such that, given $(i,j,\ell)$ in binary (where $i,j,\ell \in [\poly(n)]$) and $m \in [\ceil{\log \abs{\ringk}}]$ in binary, the algorithm outputs:
\begin{itemize}
        \item  The following information about the $i$\ts{th} gate residing in the $\ell$\ts{th} layer of $C$, denoted $g_{i,\ell}$:
            \begin{itemize}
                \item $\incoming(i,\ell,j)$: encodes the index $k$ of the gate in layer $\ell-1$ corresponding to the $j$\ts{th} incoming wire to $g_{i,\ell}$, as well as the $m$\ts{th} low-order bit of the scalar $\alpha_j \in \ringk$ attached to the wire if $g_{i,\ell}$ is linear.
                \item $\indeg(i,\ell)$: The in-degree of gate $g_{i,\ell}$.
            \end{itemize}
        The algorithm outputs $\bot$ if any of $i$, $\ell$, or $j$ are out-of-bounds.\footnote{Therefore, by convention, input gates always have all incoming wires be $\bot$.}
        \item $S_\ell^{gates}$: the number of gates in layer $\ell$ of $C$.
\end{itemize}
\end{definition}

This is defined analogously for boolean circuits.
\begin{definition}[Succinctly-uniform boolean circuits] \label{boolean-uniform-circuit} 
We say that boolean family $\{C_n\}$ is {\bf succinctly-uniform} if there is a $\poly(\log n)$-time, $O(\log n)$-space algorithm such that, given $(i,j,\ell)$ in binary (where $i,j,\ell \in [\poly(n)]$), the algorithm outputs:
\begin{itemize}
        \item  The following information about the $i$\ts{th} gate residing in the $\ell$\ts{th} layer of $C$, denoted $g_{i,\ell}$:
            \begin{itemize}
                \item $\gtype(i,\ell)$: encodes which type of gate (e.g. \texttt{AND}, \texttt{OR}, \texttt{NOT}, \texttt{THR}, etc) $g_{i,\ell}$ is.
                \item $\incoming(i,\ell,j)$: encodes the index $k$ of the gate in layer $\ell-1$ corresponding to the $j$\ts{th} incoming wire to $g_{i,\ell}$,
                as well as the binary value of incoming gate $g_{k,\ell}$ if it is a fixed constant.
                \item $\indeg(i,\ell)$: The in-degree of gate $g_{i,\ell}$.
            \end{itemize}
        The algorithm outputs $\bot$ if any of $i$, $\ell$, or $j$ are out-of-bounds.
        \item $S_\ell^{gates}$: the number of gates in layer $\ell$ of $C$.
\end{itemize}
\end{definition}

\section{Evaluating Low-Depth Circuits of Symmetric Gates}\label{sec:symcircs}

In this section, we give a fast and simple version of the circuit evaluation algorithm from \cite{bckls14}, and note several algorithmic applications of fast circuit evaluation.

\subsection{Fast SYM Circuit Evaluation}
\circeval*

\begin{proof}
By increasing the number of wires by at most a factor of $d$, we can assume that the circuit is \emph{layered}, in that all gates only read the outputs of gates on the previous layer.
We also assume without loss of generality that $w$, and all gate fan-ins, are powers of two. (Otherwise, simply add dummy gates and wires to the circuit to make this the case.)
We divide the catalytic space into ``registers'', each of which stores a value from $[w]$.

For each gate $g$ of the circuit, we assign an associated ``output register'', and $f_g$ associated ``helper registers'', where $f_g$ is $g$'s fan-in.
We will use $R_g$ and $R_g^{(1)}, \dots, R_g^{(f_g)}$ to refer to the values held by these registers, respectively, at a given point in time.
Let $\tau_g$ and $\tau_g^{(1)}, \dots, \tau_g^{(f_g)}$ denote the initial values of these registers prior to any computation.

Our goal will be to set every gate's output register to $\tau_g + b_i \mod w$, where $b_i$ is the value (0 or 1) outputted by the gate when the circuit reads its input\footnote{Observe that, if we can do so, then we can also write the answer onto an XOR-only output tape while resetting the catalytic space: First, XOR the least-significant bits of all of the final layer's output registers into the output. Then, run our algorithm. Then, do this XOR again. Finally, run our algorithm in reverse to reset all of the gates.}.
We say a gate is ``computed'' if $R_g = \tau_g + b_i \mod w$, and ``reset'' if $R_g = \tau_g$. (Note that it could also be in neither of these states.)

As in \cite{bckls14}, we do this computation recursively.
That is, we assume we have a reversible algorithm $A_\ell$ that computes all of the output registers on level $\ell$, and use this algorithm as a subroutine to yield a reversible algorithm $A_{\ell + 1}$ that computes all of the output registers on level $\ell + 1$.
In order to keep the space usage small, we ensure that we always completely erase the clean space before each subroutine call, except for a single global counter that lets the machine know which step of the algorithm to perform.
This means that the total clean space usage will be $\log(\text{\# steps})$, for the counter, plus $\max_\ell (\text{clean space overhead at the $\ell$th layer of recursion})$.

Our algorithm $A_{\ell + 1}$ proceeds as follows (erasing the clean space in between each step):
    \begin{enumerate}
        \item For each gate $g$ on layer $\ell+1$ of the circuit, determine $i_g = \sum_{g' \ | \ g' \text{ is an input gate to } g} R_{g'} \mod f_g$, and set $R_g \gets R_g - R_g^{(i_g)} \mod w$.
        \item Run $A$.
        \item For each gate $g$ on layer $\ell+1$, determine $j_g = \sum_{g' \ | \ g' \text{ is an input gate to } g} R_{g'} \mod f_g$. Now, for every value $t$ such that $g$ would output $1$ if exactly $t$ inputs were $1$, set $R_g^{j_g - t} \gets R_g^{j_g - t} + 1 \mod w$.      
        \item Run $A$ in reverse.
        \item Once more, for each gate $g$ on layer $\ell+1$, determine $i_g = \sum_{g' \ | \ g' \text{ is an input gate to } g} R_{g'} \mod f_g$. This time, set $R_g \gets R_g + R_g^{(i_g)} \mod w$.
    \end{enumerate}

    We claim this will result in every gate on the $\ell + 1$st layer being computed.
    Observe that the values of $i_g$ computed in steps i) and v) will be identical, and that the value of $j_g$ computed in step iii) will be equal to $i_g + (\text{\# of 1s among input wires to }g) \mod f_g$.
    So, step iii) will increment $R_g^{i_g}$ by $1$ if and only if the number of $1$s among inputs to $g$ would make $g$ output $1$.
    Otherwise, $R_g^{i_g}$ will remain unchanged.
    Since we subtract off from $R_g$ the initial value of $R_g^{i_g}$, and then add back in the final value of $R_g^{i_g}$, this procedure will result in $g$ being computed.
    
    Observe also that each step can be computed with low space overhead (we simply do memory lookups and compute running sums), and that this sequence of operations is reversible (every step either performs some linear operations, based on the indices specified by some registers that remain constant, or else consists of calling $A$). To bound the runtime, observe that $A_{\ell + 1}$ makes two recursive calls to $A_{\ell}$, and then for each gate performs a number of register lookups and additions proportional to that gate's fan-in, meaning at most $O(w)$ total such operations. 
    So, the overall runtime is at most $\softo{w 2^d}$.
\end{proof}

\subsection{Application: Integer Multiplication}

The ability to quickly evaluate low-depth circuits of symmetric gates is in fact a very useful algorithmic tool. Observe that \cref{thm:circeval} gives an algorithm with runtime near-linear in the circuit size whenever $d = o(\log w)$. We use this fact here to directly obtain near-linear-time catalytic logspace algorithms for several fundamental problems.

First, we observe that this gives us a near-linear-time algorithm for evaluating any $\NC^1$ circuit:

\begin{lemma}\label{lem:nc1}
    Given a fan-in $2$ boolean circuit of size $S$ and depth $O(\log(S))$, there exists a catalytic algorithm which computes the circuit's output using $O(\log(S))$ clean workspace and $S^{1 + O(1/\log\log(S))}$ time.
\end{lemma}
\begin{proof}
    We observe that any such circuit can be easily converted into a SYM circuit of depth $\log(S)/\log\log(S)$ and size $S^{1 + O(1/\log\log(S))}$.
    That is, we show how to replace $\log\log(S)/100$ layers of fan-in-$2$ gates with $O(1)$ layers of arbitrary fan-in SYM gates. 
    Observe that an output at the end of those $\log\log(S)/100$ layers can only depend on at most $2^{\log\log(S)/100} = \log(S)^{1/100}$ inputs.
    We can build a constant-depth SYM circuit which checks whether those inputs equal each of the $2^{\log(S)^{1/100}}$ possible settings, and returns the correct value for each.
    This requires $2^{O(\log(S)^{1/100})}$ wires.
    So, using $2^{O(\log(S)^{1/100})}$ wires for each gate of the original circuit, we can replace every $\log\log(S)/100$ layers of the original circuit with circuits of this form.
    The overall statement holds by plugging that resulting circuit into \cref{thm:circeval}.
\end{proof}

From this, we immediately get fast catalytic algorithms for any problems for which we know small $\NC^1$ circuits. For instance:

\begin{corollary}\label{cor:intmult}
    There exists a catalytic algorithm using $O(\log(n))$ clean space and $n^{1 + o(1)}$ time which takes as input a pair of $n$-bit integers and computes their product.
\end{corollary}
\begin{proof}
    The classic Sch{\"o}nhage--Strassen algorithm for integer multiplication gives a (succinctly uniform) $\NC^1$ circuit of size $O(n\log(n)\log\log(n))$~\cite{SchonhageStrassen1971, Wegener1987}. Applying \cref{lem:nc1} to this circuit yields the result.
\end{proof}

Recall that Sch{\"o}nhage--Strassen works by computing a discrete Fourier transform over $\Z/(2^m + 1)\Z$ --- the same argument as above shows that such DFTs can be performed in near-linear time catalytically.
As we observe in \cref{sec:lincircs}, in fact there exists an algorithm with the even faster runtime $\widetilde O(n)$, over any ring with efficiently-computable operations, which in turn also gives $\widetilde O(n)$-time polynomial multiplication.

\subsection{Application: Sorting}

We do not know small $\NC^1$ circuits for sorting lists of integers, but fortunately, if we are somewhat more careful, we can nonetheless also apply \cref{thm:circeval} here:

\begin{corollary}
    There exists a catalytic algorithm using $O(\log(n + b))$ clean space and $n^{1 + o(1)} b$ time which takes as input a length-$n$ list of $b$-bit numbers, and computes the sorted list.
\end{corollary}
\begin{proof}
    Recall that there exists an $O(\log(n))$-depth \emph{sorting network}~\cite{Ajtai1983Sorting}.
    A sorting network operates over $n$ wires, and consists of ``comparators'' which swap the values of two wires if they are out of sorted order.

    If this network was converted into a fan-in 2 Boolean circuit, each comparator would require depth $\log(b)$, so the overall circuit would have depth $O(\log(n)\log(b))$, which is much too high to apply \cref{lem:nc1}.
    However, we can nonetheless produce a $o(\log(n))$-depth, $n^{1+o(1)}b$-size SYM circuit, by a similar approach to the proof of that lemma. That is, we show how to evaluate $\log\log(n)/100$ layers of a sorting network using a constant-depth SYM circuit with $O(bn^{1+1/\log\log(n)})$ wires.

    Suppose we have written down the values on all of the wires prior to this sequence of layers, and want to compute the value of one of the wires after the sequence of layers.
    This wire's output depends on at most $2^{\log\log(n)/100} = \log^{1/100}(n)$ of the initial values.
    Observe that there are only $\log^{1/100}(n)! \leq 2^{\log^{1/50}(n)}$ possible orderings of those $\log^{1/100}(n)$ many values, and that the ordering of those values is sufficient to determine which of them should be written to the output wire.
    Now, for a given candidate ordering, there exists a $O(b 2^{\log^{1/50}(n)})$-wire $\AC^0$ circuit which evaluates whether the values are thus ordered: we can check whether a given pair of values is in the correct order in $O(b)$ size and depth $O(1)$~\cite{Chandra1984Constant}, so we simply do all of these checks in parallel and take their AND. So, to compute the $i$th bit on output value of the wire, we can simply compute in parallel for each initial value ordering whether that ordering is correct, and then take the OR over all orderings of ``this initial value ordering is correct AND the $i$th bit of the value which would get output under this initial value ordering is a $1$''.

    This overall gives a SYM circuit for sorting with depth $O(\log(n)/\log\log(n))$ and size $O(n\log(n) \cdot b 2^{\log^{1/50}(n)}) = O(b n^{1+o(1)})$. Since the AKS sorting network is DLOGTIME uniform, and the identity of a gate or wire in the above constructed circuit can be computed in logarithmic space and $n^{o(1)}$ time given the original sorting network, we can apply \cref{thm:circeval} to evaluate it in time $bn^{1+o(1)}$.
\end{proof}

We observe that by composing catalytic algorithms one can use these results to get fast algorithms for decision (and counting) problems:
 \ksum*
\begin{proof}
    The standard reduction from {\sc kSUM} to {\sc 2SUM} is efficiently computable, in the sense that a given bit of the {\sc kSUM} instance can be computed in $\log(n)$ space and $\polylog(n)$ time given read access to the {\sc 2SUM} instance~\cite{Erickson1995}.
    So, it suffices to give an $n^{1 + o(1)}$-time, $O(\log n)$ clean space, catalytic algorithm for {\sc 2SUM}.
    This follows from the standard reduction from {\sc 2SUM} to sorting, which can be implemented catalytically.
    
    That is, take one of the lists, negate all of the elements in it, add some small $\eps$ to all of them, and merge it with the other list.
    Then, use the $n^{1+o(1)}$-time $\CL$ algorithm for sorting.
    Finally, observe that there is a near-linear-size $\TC^0$ circuit which returns whether any adjacent elements of a list differ by exactly $\eps$ --- this yields a near-linear-time $\CL$ algorithm by \cref{thm:circeval}, which reads its input via XORing into the catalytic tape a constant number of times, and hence can be composed with the previous algorithm.
    The composition of these algorithms will decide {\sc 2SUM}.    
\end{proof}

We note that a similar idea\footnote{One can similarly reduce to \#{\sc 2SUM}, and begin by sorting, but now in order to construct a near-linear-size, $o(\log(n))$-depth threshold circuit which performs the relevant counting step, one needs to be somewhat more careful. One can get this construction e.g. by breaking up into blocks of size $n^{1/\log\log(n)}$, spending $n^{2/\log\log(n)}$ time counting in each block, and recursing $\log\log(n)$ times, as Allender and Kouck\'{y} do for hardness magnification~\cite{AllenderKoucky2010}.} will also allow computing \#{\sc kSUM} in $n^{\ceil{k/2}+o(1)}$-time $\CL$. A modification of this \#{\sc 2SUM} construction, instead taking the $k$th power of the element count in equal-value blocks, gives:
\begin{restatable}{corollary}{freqmoments} \label{cor:freqmoments}
Let $A$ be a list of $n^{o(1)}$-bit integers. Let $n=\abs{A}$ and let $k>1$ be a constant.
There exists a catalytic algorithm computing the $k$\ts{th} frequency moment
\[
F_k=\sum_{i \in \Z} (f_i)^k,
\]
where $f_i$ is the frequency of value $i$ in $A$. This algorithm takes $n^{1+o(1)}$ time, $n^{1+o(1)}$ catalytic space, and $O(\log n)$ clean space.
\end{restatable}
We note that while \cite{datastream2026,streaming26} show there exist $\widetilde O(n)$ time catalytic algorithms computing frequency moments, their algorithms would use catalytic space proportional to the size of the universe (i.e. they would only get a bound of $2^{n^{o(1)}}$ catalytic space).

\section{Evaluating Low-Degree Arithmetic Circuits}\label{sec:lincircs}
In this section, we observe that with some modifications the original \cite{bckls14} approach for circuit evaluation will actually allow fast evaluation of even high-depth arithmetic circuits, so long as most layers consist only of linear operations. This allows us to obtain fast catalytic algorithms for Fourier transforms, matrix multiplication, and applications thereof.

\subsection{Fast Low-Multiplicative-Depth Circuit Evaluation}

\begin{definition}
    For a ring $\calK$, an $\#\NC(\calK)$ circuit has the following gateset:
    \begin{enumerate}
        \item fan-in 2 addition gates,
        \item fan-in 2 product gates, and
        \item fan-in $1$ ``$\times x$'' and ``$+ x$'' gates for each constant $x \in \calK$.
    \end{enumerate}
    We say that such a circuit has \defn{multiplicative depth} $d$ if its gates are presented grouped into layers such that gates in each layer depend only on gates in previous layers, and such that at most $d$ of those layers involve (fan-in 2) product gates.
\end{definition}

We show that we can evaluate succinctly uniform $\#\NC(\calK)$ circuits in near-linear time so long as they have sublogarithmic multiplicative depth, and so long as the ring operations are efficiently computable.
Specifically, the ring requirements are as follows, following \cite[Section 4.1]{bckls14} but with a further time constraint:

\begin{definition}[\cite{bckls14}, Section 4.1] \label{def:uniform-rings}
    Let $\ringk=\ringk_n$ be a family of rings. We say $\calK$ is \defn{$(S,T)$-uniform} if there exists a bijection $e_n$ mapping $\ringk_n$ to the lexicographically first $\abs{\ringk}$ strings of $\set{0,1}^{\ceil{\log \abs{\ringk_n}}}$, such that there exist $S(n)$-space, $T(n)$-time algorithms for the following operations:
    \begin{itemize}
        \item On read-only input $(1^n,e(u),e(v))$, output (each bit of) $e(u \circ v)$, for operation $\circ \in \set{+,-,\times}$.
        \item On read-only input $(1^n,e(v))$ and $e(u)$ on a read-write tape, transform $e(u)$ in place into $e(u+v)$.
        \item On input $(1^n)$, output (each bit of) $e(-1),e(0),e(1)$ and $\abs{\ringk_n}$.
    \end{itemize}
\end{definition}
Note that, for instance, $\{\Z/m_n\Z\}_{n\in\N}$ is $(O(\log \log m), \polylog(m))$-uniform for any uniform sequence $\{m_n\}_{n\in\N}$.
We have the following:

\begin{theorem}\label{thm:arithcirceval}
    Let $\calK$ be an $(S,T)$-uniform ring.
    There exists a catalytic algorithm using time $\softo{w 3^d T}$, clean space $O(\log(w) + d + \log(T) + S)$, and dirty space $\softo{w \log|\calK_n|}$, which takes in a $\#\NC(\calK_n)$ circuit with $w$ wires and multiplicative depth $d$, and a setting of all of the input wires to the circuit, and computes the value of all of the circuit's output wires.
\end{theorem}
\begin{proof}
    The proof is essentially identical to that of \cite{bckls14}, except that we make a distinction between different types of multiplication gates (linear and non-linear) and adjust accordingly.
    We think of the dirty space as consisting of a sequence of ``registers'', each holding a value over $\calK_n$.
    Note that, if $|\calK_n|$ is a power of $2$, this is trivial to do, but otherwise it is non-obvious one can treat the dirty space this way.
    Nonetheless, \cite{chmel2026frontier} shows how, at polylogarithmic overhead in time and dirty space, one can assume without loss of generality that one has access to such registers over $\calK_n$. (This follows by compressing the dirty space whenever a large block of tape cells contains no value in the appropriate range --- see their Appendix A. Their algorithm requires a pre-processing and post-processing phase, so we assume our entire procedure below happens between these.)

    Let $\tau_i \in \calK_n$ denote the initial contents of the $i$th register, and $R_i$ denote the value currently held at a given point in the algorithm.
    As in \cref{thm:circeval}, we say that the $i$th gate's value is ``reset'' if $R_i = \tau_i$, and ``computed'' if $R_i = \tau_i + v_i$, where $v_i$ is that gate's output in the arithmetic circuit. 
    Assume we have a reversible algorithm $A_\ell$ that results in all of the gates up to and including the $\ell$th multiplicative layer being converted from ``reset'' to ``computed'', and leaves the clean space empty except for a single global time counter.
    We give a reversible algorithm $A_{\ell + 1}$ which makes $3$ calls to $A_\ell$ and takes $\softo{w T}$ time.
    This algorithm will require further helper registers $R_i^{(1)}$ and $R_i^{(2)}$ for every $i$ corresponding to a product gate.

    The algorithm proceeds as follows (erasing the clean space after each step):
    \begin{enumerate}
        \item For each gate $i$ (in order of layers, from the output layer backwards): 
        \begin{itemize}
            \item If $i$ is an addition gate in-between the $\ell$th and $\ell+1$st multiplicative layers, set $R_i \gets R_i - R_j - R_k$, where $j$ and $k$ are the two input gates.
            \item If $i$ is a $\times x$ gate in-between the $\ell$th and $\ell+1$st multiplicative layers, set $R_i \gets R_i - R_j \times x$, where $j$ is the input gate. 
            \item If $i$ is a $+ x$ gate in-between the $\ell$th and $\ell+1$st multiplicative layers, set $R_i \gets R_i - R_j$, where $j$ is the input gate. 
            \item If $i$ is a product gate on the $\ell + 1$st layer of the circuit, set $R_i \gets R_i + R_jR_k + R_j R_i^{(1)} + R_i^{(2)} R_k$, where $j$ and $k$ are the two input gates.
        \end{itemize}
        \item Run $A_\ell$.
        \item For each gate $i$ (in order of layers, from the input layer forwards): 
        \begin{itemize}
            \item If $i$ is an addition gate in-between the $\ell$th and $\ell+1$st multiplicative layers, set $R_i \gets R_i + R_j + R_k$, where $j$ and $k$ are the two input gates.
            \item If $i$ is a $\times x$ gate in-between the $\ell$th and $\ell+1$st multiplicative layers, set $R_i \gets R_i + R_j \times x$, where $j$ is the input gate. 
            \item If $i$ is a $+ x$ gate in-between the $\ell$th and $\ell+1$st multiplicative layers, set $R_i \gets R_i  + R_j + x$, where $j$ is the input gate. 
            \item If $i$ is a product gate on the $\ell + 1$st layer of the circuit, set $R_i\gets R_i + R_jR_k$, set $R_i^{(1)} \gets R_i^{(1)} + R_k$, and set $R_i^{(2)} \gets R_i^{(2)} + R_j$, where $j$ and $k$ are the two input gates.
        \end{itemize}
        \item For each gate $i$ (in order of layers, from the output layer backwards): 
        \begin{itemize}
            \item If $i$ is an addition gate in-between the $\ell$th and $\ell+1$st multiplicative layers, set $R_i \gets R_i - R_j - R_k$, where $j$ and $k$ are the two input gates.
            \item If $i$ is a $\times x$ gate in-between the $\ell$th and $\ell+1$st multiplicative layers, set $R_i \gets R_i - R_j \times x$, where $j$ is the input gate. 
            \item If $i$ is a $+ x$ gate in-between the $\ell$th and $\ell+1$st multiplicative layers, set $R_i \gets R_i - R_j - x$, where $j$ is the input gate. 
        \end{itemize}
        \item Run $A_\ell$ in reverse.
        \item For each gate $i$ (in order of layers, from the input layer forwards): 
        \begin{itemize}
            \item If $i$ is an addition gate in-between the $\ell$th and $\ell+1$st multiplicative layers, set $R_i \gets R_i + R_j + R_k$, where $j$ and $k$ are the two input gates.
            \item If $i$ is a $\times x$ gate in-between the $\ell$th and $\ell+1$st multiplicative layers, set $R_i \gets R_i + R_j \times x$, where $j$ is the input gate. 
            \item If $i$ is a $+ x$ gate in-between the $\ell$th and $\ell+1$st multiplicative layers, set $R_i \gets R_i + R_j$, where $j$ is the input gate. 
            \item If $i$ is a product gate on the $\ell + 1$st layer of the circuit, set $R_i \gets R_i - R_j R_i^{(1)} -  R_i^{(2)} R_k$, where $j$ and $k$ are the two input gates.
        \end{itemize}
    \end{enumerate}

    Observe that, for any non-product gate $i$ between the $\ell$th and $\ell+1$st layers, $i$ will be reset initially, computed after step iii), and then reset again after step vi). By the same analysis as in \cite{bckls14}, all of the product gates will therefore end up computed at the end of this program. 
    All of the steps are reversible, so we now have a reversible algorithm $A_{\ell + 1}$, which makes two calls to $A_\ell$ plus an additional $\softo{wT}$ runtime overhead, and results in all of the gates on multiplicative layer $\ell+1$ being computed.
    Note that this algorithm by default wouldn't result in the gates prior to multiplicative layer $\ell + 1$ being computed. However, we can easily fix this by running $A_\ell$ again. If the final tail of layers in the circuit are all linear (the last layer isn't multiplicative), we can compute up to those layers by simply running steps i), $A_\ell$ (where $\ell$ is the last multiplicative layer), and iii) on them.

    Now, evaluating $A_d$ will take runtime $\softo{3^d w T}$, and clean space $O(\log(w) + d + \log(T))$ (for the global time counter) plus $O(S)$ (for the individual register operations).
    $A_d$ results in the circuit's outputs being added into the dirty registers as elements of $\calK_n$ --- we observe in \cref{appendix:registers} that such a program can efficiently be converted into one that instead bitwise-XORs the circuit outputs into the dirty space.
\end{proof}

\subsection{Application: Fourier Transforms}

One particularly useful corollary of \cref{thm:arithcirceval} is the fact that we can implement fast Fourier transforms catalytically:

\begin{definition}[Principal Root of Unity] \cite[p.11]{Bini1994-hg} \label{principal-rou-def}
    Let $\ringk$ be a finite ring and $1< m <\abs{\ringk}$.
    $\omega_m \in \ringk$ is a \emph{principal root of unity} of order $m$ iff:
    \begin{itemize}
        \item $(\omega_m)^m=1$.
        \item $\sum_{j=0}^{m-1} (\omega_m)^{ij}=0$ \text{ for all } $1 \leq i < m$.
    \end{itemize}
\end{definition}

\begin{definition}[Discrete Fourier Transform, DFT] \label{dft-def}
Let $\vecx=(x_0,\cdots,x_{n-1})^T \in \ringk^{n}$ be a column vector over the ring $\ringk$.
Define the matrix
\[
F_n=\begin{pmatrix}
    1 & 1 & 1 & \cdots & 1\\
    1 & \omega_n^1 & \omega_n^2 & \cdots & \omega_n^{n-1}\\
    \vdots & \ddots & \vdots\\
    1 & \omega_n^{1(n-1)} & \omega_n^{2(n-1)} & \cdots & \omega_n^{(n-1)(n-1)}\\
\end{pmatrix}
\]
in $\ringk^{n \times n}$ where $\omega_n$ is an $n$\ts{th} order principal root of unity in $\ringk$.
The discrete Fourier transform is the linear transformation $\vecx \mapsto F_n \cdot \vecx$.
\end{definition}

\begin{corollary}
    Over any $(S,T)$-uniform ring $\calK=\ringk_n$, for $n$ a power of $2$, if $\calK$ admits a principal $n$th root of unity, with both that root and its inverse computable in $S$ space and $T$ time, then there exists a catalytic algorithm running in time $\softo{n T}$, clean space $O(\log(n) + \log(T) + S)$, and dirty space $\softo{n \log|\calK|}$, which takes as input a sequence $x \in \calK^n$, and outputs the discrete Fourier transform $F_nx$.
\end{corollary}
\begin{proof}
    The classic Number Theoretic Transform algorithm and its inverse are directly implementable as $\#\NC(\calK)$ circuits of multiplicative depth $0$ --- i.e., with only addition and scalar multiplication gates --- assuming a principal $n$th root of unity and its inverse are available as constants~\cite[Section 30.2]{CLRS}. \cref{thm:arithcirceval} therefore allows such circuits to be evaluated catalytically with only polylogarithmic overhead in time complexity.
\end{proof}

We could have obtained a similar result in \cref{sec:symcircs}, but recall that there our algorithm would have required $n^{1 + o(1)}$ time, as opposed to $\softo{n}$, and would have been difficult to make work over as general a class of rings, since we would have to implement the ring operations themselves as sufficiently low-depth symmetric circuits.

The ability to compute fast Fourier transforms also allows us to obtain quasilinear-time catalytic algorithms for a number of applications. For instance:

\begin{corollary}\label{cor:polymult}
    For any $(S,T)$-uniform ring $\calK$, there exists a catalytic algorithm which takes as input two degree-$n$ polynomials $p, q \in \calK[x]$ and outputs their product $pq$, using $\softo{nT}$ time, $O(\log(n) + S)$ clean space, and $\softo{nT}$ dirty space.
\end{corollary}
\begin{proof}
    The quasilinear-time algorithm of Cantor and Kaltofen is bilinear, meaning it can be implemented by a $\#\NC(\calK)$ circuit of multiplicative depth $1$~\cite{CantorKaltofen91}.
    (Note that this algorithm does not even require knowing roots of unity.)
    Thus, \cref{thm:arithcirceval} again yields a quasilinear-time catalytic algorithm.
\end{proof}

\begin{corollary}[\#3SUM for Small Universes]\label{cor:3sumsmall}
Let $A,B,C \subseteq [0,M]$ be three integer lists for $2\leq M\leq \poly(n)$, let $t$ be an integer, and let $n=\max\set{\abs{A}, \abs{B},\abs{C}}$.
There is a catalytic algorithm counting the number of triplets $a\in A$, $b \in B$, and $c \in C$ such that $a+b+c=t$ using $\softo{n+M}$ time, $\softo{M}$ dirty space, and $O(\log n)$ clean space.
\end{corollary}
\begin{proof}
    As is standard (see, e.g., \cite[Ex. 30.1-7]{CLRS}), we can represent $A$, $B$, and $C$ by degree-$M$ polynomials $a$, $b$, and $c$, respectively, where the coefficient on $x^i$ is equal to the number of times that $i$ appears in the set. (We only need to sequentially iterate over each $A,B,C$ once to add these polynomials to the appropriate registers. Given each $A_i$ (resp. $B_i$, $C_i$) value, add 1 into the corresponding coefficient register.)
    By two applications of the construction from \cref{cor:polymult}, we can catalytically compute the product of all three of these polynomials over $\Z/((M+1)n^3+1)\Z$ in $\softo{n+M}$ time, $O(M\log M \log n)$ dirty space, and $O(\log n)$ clean space.
    Now, by computing the coefficient on $x^t$ in this product directly into the clean space as opposed to a catalytic register, we obtain the desired result.
\end{proof}

We can also apply our algorithm to obtain (static) string pattern matching with wildcards. Let $\Sigma=\Sigma' \cup \set{\star}$ be the alphabet. Given a pattern $p$ over $\Sigma$ and a sub-string $s'$ over $\Sigma$, we say $s'$ matches $p$ if on every index $i$ either $p_i=\star$, $s_i'=\star$, or $p_i=s_i'$. Assume, WLOG, that $\abs{\Sigma} \leq (n+m)+1$ (including wildcard symbol $\star$) and $m\leq n$ for length-$m$ pattern $p$ and length-$n$ string $s$.

\begin{corollary}[Wildcard String Matching]\label{cor:stringmatching}
    Assume strings $s \in \Sigma^n$ and $p \in \Sigma^m$ are provided as input.
    There exists a catalytic algorithm using time $\softo{n}$, $O(n \log^2 n)$ dirty space, and $O(\log n)$ clean space computing an indicator vector $v$ where $v_i=1$ iff pattern $p$ matches substring $s_{i:i+m-1}$.
\end{corollary}
\begin{proof}
    The alphabet symbols are provided as positive integers of magnitude at most $n+m$.
    Replace any $\star$ in $p$ and $s$ with zeros.
    An algorithm of \cite{clifford2007simple} proceeds by computing, for each $i$, the quantity
    \begin{gather*}
        q_i=\sum_{j=0}^{m-1} p_js_{i+j} \left( p_j - s_{i+j} \right)^2=\\
        \sum_{j=0}^{m-1} p_j^3s_{i+j} - 2p_j^2s_{i+j}^2 + p_js_{i+j}^3.
    \end{gather*}
    This quantity $q_i$ is 0 if pattern $p$ matches substring $s_{i:i+m-1}$ and positive otherwise.
    
    The squared and cubed values in the quantities can be computed by an $O(n)$-size circuit with multiplicative depth 2.
    The individual terms can then be computed and accumulated using polynomial multiplication over a sufficiently large ring $\Z/c\Z$ ($c=\poly(n)$) to avoid modular overflow. Overall, this is a multiplicative depth 3 arithmetic circuit. Therefore, by \Cref{thm:arithcirceval} along with the construction of \Cref{cor:polymult}, we can compute this quantity $q_i$ for all $1 \leq i \leq n-m+1$ in time $n \cdot \poly(\log n)$ catalytically into registers.
    Finally, convert this modular ring register vector output to XOR output form by \Cref{lem:mod-to-xor}. Then, apply \Cref{thm:circeval} to a small depth fan-in 2 circuit (size $(n-m+1) \cdot \poly(\log n)$, depth $O(\log \log n)$) over each register value which outputs 1 if $q_i=0$ and outputs 0 otherwise. In total, this uses $n \cdot \poly(\log n)$ time and $O(n \log n) \cdot O(\log c)$ dirty space.
\end{proof}

\subsection{Application: Matrix Multiplication} \label{sec:mm}

Another problem with nontrivial algorithms of low multiplicative depth is matrix multiplication.
Here, even though we don't know the optimal runtime, we know that it is achievable catalytically. 
We observe the following:

\begin{definition}
The matrix multiplication exponent $\omega$ is the infimum over all $r$ such that $n\times n$ matrices over $\mathbb{C}$ can be multiplied by arithmetic circuits\footnote{\emph{Non-uniform} arithmetic circuits with allowed division gates.} of size $O(n^r)$.
\end{definition}

\begin{corollary}\label{cor:mm}
There is a catalytic algorithm which, given $m\geq 2$ and $A,B\in(\mathbb{Z}/m\mathbb{Z})^{n\times n}$, computes $AB$ using clean space $O(\log n+\log m)$, time $n^{\omega+o(1)}\polylog(m)$, and dirty space $n^{\omega+o(1)}\polylog(m)$.
\end{corollary}
\begin{proof}
By standard results from the matrix multiplication literature \cite{BCS97, DBLP:journals/siamcomp/Schonhage81}, for every $\eps>0$ there is a bilinear algorithm over $\Q$ that multiplies $k\times k$ matrices for some $k$ using at most $k^{\omega+\eps}$ products.\footnote{It is known that the exponent of matrix multiplication (over fields) can only depend on the characteristic of the field, if at all~\cite{DBLP:journals/siamcomp/Schonhage81}.}
By doing a brute-force search over all such bilinear algorithms with small (e.g. at most $\log\log\log(n)$) size and bit complexity, we can therefore find one using at most $k^{\omega + o(1)}$ products for some $k$.\footnote{Instead of iterating over small arithmetic circuits, we can also simply brute force the coefficients witnessing a rank upper bound of the $\langle k, k, k \rangle$ matrix multiplication tensor rank. (The arithmetic matrix exponent and the infimum of the logarithm base $k$ of this ``tensor rank'' are \emph{equivalent} over characteristic zero fields. See e.g. \cite{Strassen73}.) Using this method, brute forcing over the possible rank bounds for $k \leq \log \log \log n$ and bit complexity (of the $\Q$ coefficients) $\log \log \log n$ is sufficient for our purposes to achieve $\omega+o(1)$. Observe this is stronger than $\omega+\eps$ for all $\eps>0$ as, in that case, the algorithm's leading constant would depend on $\eps$.}
This is a bilinear algorithm, and hence an $\#\NC(\Q)$ circuit of multiplicative depth $1$.
Applying this bilinear algorithm recursively, and then unrolling the recursion, gives an $\#\NC(\Q)$ circuit of multiplicative depth $1$ and size $n^{\omega + o(1)}$ for $n\times n$ matrix multiplication.

In order to use a ring we can actually store in registers, we let $d$ be the least common multiple of all the denominators used in that bilinear algorithm, and consider working in $\mathbb{Z}/m'\mathbb{Z}$, where $m' = dnm^2 + 1$ --- here, we know $d^{-1} \equiv -nm^2$, so we can easily replace the constants over $\Q$ with constants over $\Z/m'\Z$.
This gives a fast catalytic algorithm for matrix multiplication over $\Z/m'\Z$; in order to multiply over $\Z/m\Z$, we first take the representative of each number in $[m]$, think of those as elements of $\Z/m'\Z$, and do the multiplication there. All of the entries of the product lie in $[1, nm^2]$, so the results $\mod m$ are uniquely determined by the results $\mod m'$.
So, this algorithm adds the desired output values into dirty registers mod $m'$. In \cref{appendix:registers}, we observe that it is possible to generically convert such an algorithm into one which bitwise-XORs the desired output values mod $m$ into the dirty space.
\end{proof}

We can then apply this to get fast catalytic versions of algorithms involving matrix multiplication. For instance, using a reduction due to Itai and Rodeh \cite{itairodeh77},

\triangle*
\begin{proof}
    Consider an arithmetic circuit which cubes the adjacency matrix and then takes the trace. 
    Since matrix multiplication can be done in multiplicative depth $1$ and size $n^{\omega + o(1)}$, and trace can be done in multiplicative depth $0$ and size $O(n)$, overall this circuit has multiplicative depth $2$ and size $n^{\omega + o(1)}$.
    We can therefore catalytically compute the result applying the construction of \Cref{cor:mm} to get the result over $\Z/(n^{3}+1)\Z$. 
    The result will be $6$ times the number of triangles in the graph; since the value is $O(\log(n))$ bits, we can compute it directly into the clean space and then divide by $6$ at the end.
\end{proof}

By a classic reduction due to Nešetřil and Poljak~\cite{NP85}, this similarly allows us to count $3k$-cliques in time $n^{k\omega + o(1)}$ for constant $k$. Their reduction is local (i.e. given read-access to the graph $G$, we can find the ith bit of the input graph $G'$ to the triangle counting algorithm in logarithmic space and polylogarithmic time), which allows us to apply \Cref{thm:triangle} directly.\footnote{A technical detail is that, to ensure efficient locality, we must use a slightly modified reduction which  produces a $n^k$-node graph: each $k$-tuple of values in $[n]$ is a node, and an edge $(u,v)$ exists if both $u$ and $v$ are in strictly increasing order and form a $2k$ clique. The number of triangles equals $c=(3k)!/(k!^3\cdot 3!)$ times the number of cliques in $G$. We can simply compute the clique count into our clean space first and then divide by $c$.} Given a counting algorithm, we can also \emph{find} a $3k$-clique catalytically, using a space-efficient version of a standard search-to-decision reduction (\cite{DBLP:journals/jacm/WilliamsW18}). This gives:

\begin{corollary}[Clique Counting] \label{cor:cliques}
    Let $G$ be an $n$-vertex simple undirected graph, given as an adjacency matrix. Let $k$ be a constant positive integer.
    There exists a catalytic algorithm which, using $n^{k\omega+o(1)}$ time, $n^{k\omega +o(1)}$ dirty space, and $O(\log n)$ clean space, counts the number of $3k$-cliques in $G$ (and outputs a clique if one exists).
\end{corollary}

Using the color-coding approach of \cite{alon1994color}, we can detect $k$ paths (and $k$ cycles) in directed graphs in matrix multiplication time with an exponential dependency on $k$. We first solve an auxiliary problem. Given a set of $t$ colors $C$ such that each node in the graph is labeled with such a color, a $(t-1)$-length path $p$ is said to be \textbf{$C$-colorful} if each color in $C$ appears exactly once in $p$.
\begin{lemma}[Colorful Paths] \label{lem:colorful}
    Let $G=(V,E)$ be an $n$-vertex simple directed graph given as an adjacency matrix.
    Let $C \subseteq [n]$ be a set of $t$ colors. Suppose the colors of each vertex of $G$ are provided as input. There exists a catalytic algorithm which detects if a $C$-colorful path exists in $G$ in $n^{\omega+o(1)}2^{O(t)}$ time, $(t \log n) \cdot n^{\omega+o(1)}$ dirty space, and $O(t \log t + \log n)$ clean space.
\end{lemma}
\begin{proof}
    We first detail the recursive algorithm of \cite{alon1994color}. The goal will be to recursively compute a matrix $D$ such that $D_{ij}>0$ if there exists a $C$-colorful path starting at $i$ and ending at $j$ in $G$, and $D_{ij}=0$ otherwise. (The final answer is simply the sum over all $D_{ij}$.) Fix a partition $C=C_1 \sqcup C_2$, and let $V_{C_1}$ and $V_{C_2}$ be the vertices in $G$ with those respective colors. 
    Recursively compute matrix $A_{C_1}$ which encodes the above colorful path pair relation for vertices in $V_1$ and color set $C_1$. Similarly, compute matrix $A_{C_2}$ for $V_2$ with colors $C_2$. Define $(B_{C_1,C_2})_{ij}$ to be 1 iff $i \in V_1, j \in V_2$ and $(i,j) \in E$ (and 0 otherwise), and compute  $A_{C_1}\cdot B_{C_1,C_2} \cdot A_{C_2}$. (Each entry of matrix $B$ can be computed in $O(\log n)$ clean space and $\polylog n$ time using the adjacency matrix.) Accumulating a sum of this quantity over all partitions $C_1$ and $C_2$ of $C$ of sizes $\ceil{t/2}$ and $\floor{t/2}$ respectively in a fixed ordering yields the desired matrix $D$. There are at most $\binom{t}{\ceil{t/2}} \leq 2^t$ such partitions. (As a base case, when $\abs{C}=1$, return the identity matrix.) All matrices remain of dimension $n$ throughout.

    To encode this as a low multiplicative depth circuit, first observe that the recursion depth is $\log t + O(1)$. We can make all the calls at some depth to appear in parallel within the circuit. Each section within a level (provided color set $C'$) consists of getting $D$ by computing $A_{C_1'}\cdot B_{C_1',C_2'} \cdot A_{C_2'}$ (given $A_{C_1'},A_{C_2'}$ from the level below) using fast matrix multiplication and then summing over all $C'=C_1' \sqcup C_2'$ partitions. This will have multiplicative depth $O(\log t)$ and size $2^{O(t)}\cdot n^{\omega+o(1)}$.
     
    We now resolve some details. Let $s_i'=(t/2^i+1)$. At the $i$th level (0th is initial call), we have $2^{s_i}$ such sections, for $s_i=\smash{\sum_{j=0}^{i-1} s_{j}'}$. Each section at level $i$ invokes $2^{s_i'}$ new recursive instances. (We round up so the number of calls is always this power of two. WLOG, let the rest of the calls just return the zero matrix.) In order to figure out which color set $C'$ is the input to a given section within a level, we can first figure out an (at most) $\log t$-length tuple where the $i$th entry indicates the index in $[2^{s_i'}]$ of which recursive path we took in each layer. Using this tuple find the appropriate color set by simulating which color partitions were recursively chosen. This incurs a multiplicative factor of $2^{O(t)}$ and only $O(t \log t)$ additional clean space is needed.
    By applying the construction of \Cref{cor:mm} and using \Cref{thm:arithcirceval} over an appropriate modular ring we obtain a $n^{\omega+o(1)}2^{O(t)}$ time catalytic algorithm using $O(t \log t + \log n)$. Choosing a $\Theta(t \log n)$-bit modular ring is sufficient to avoid overflow, so the dirty space usage is $(t \log n) \cdot n^{\omega+o(1)}$.
\end{proof}

Now, following the approach of \cite{alon1994color}, we can detect $k$-paths given a colorful path detector using perfect hashing:
\begin{restatable}[Detecting Paths]{corollary}{paths} \label{cor:paths}
    Let $G$ be an $n$-vertex simple directed graph given as an adjacency matrix. Let $k>2$ be a positive integer.
    There exists a catalytic algorithm deciding if a $k$-length simple path in $G$ exists in $2^{O(k)}\cdot n^{\omega+o(1)}$ time, $(k \log n) \cdot n^{\omega+o(1)}$ dirty space, and $O(k\log k + \log n)$ clean space.
\end{restatable}
\begin{proof}
    Use the $(k+1)$-perfect hash family from $[n]$ to $[k+1]$ of \Cref{perfect-hash}, and iterate over all $2^{O(k)}\poly(\log n)$ hash functions using $O(k + \log n)$ space.
    For each hash function $h$, let a node $j$ be colored with $h(j) \in C=[k+1]$. This induces a coloring on the vertices, which is computable given any vertex in $\polylog(n)$ time and $O(\log n)$ clean space.
    For any simple length-$k$ path $p=(p_1,\cdots,p_{k+1})$ in $G$, there will be a hash function in the family such that $p$ is $C$-colorful, using the $(k+1)$-perfect property (\Cref{perfect-hash-def}). Therefore, we can simply take the \texttt{OR} of the outputs of the algorithm from \Cref{lem:colorful}.
    \qedhere
\end{proof}

By proceeding almost exactly as the $k$-length path approach above, modifying it slightly, we can obtain an algorithm detecting $k$-length \emph{cycles}. The idea is that if the possible start and end vertices of colorful paths of length exactly $(k-1)$ are appropriately encoded in the final matrix (before summing all entries) of \Cref{lem:colorful}, we can then filter for paths whose end vertex has an edge directed towards the start vertex via an additional matrix multiplication --- this forms a $k$-cycle. Iterating over hashes in perfect families exactly as in \Cref{cor:paths} yields:
\begin{restatable}[Detecting Cycles]{corollary}{cycles} \label{cor:cycles}
    Let $G$ be an $n$-vertex simple directed graph, and let $k>2$ given as an adjacency matrix $A$.
    There exists a catalytic algorithm deciding if a $k$-length simple cycle in $G$ exists in $2^{O(k)}\cdot n^{\omega+o(1)}$ time, $(k \log n + k) \cdot n^{\omega+o(1)}$ dirty space, and $O(k\log k + \log n)$ clean space. 
\end{restatable}

We can use $O(k \log n)$ bits of additional clean space, with no asymptotic time or catalytic space loss, to \emph{find} the $k$-length cycle or path, not just detect it.\footnote{As we iterate over each of the hash functions, stop when it is found that a cycle (resp. path) exists, and store the corresponding hash function. At least one of the cycles (resp. paths) of $G$ will be colorful. Now, we follow essentially the same approach as the cliques search to decision reduction. Brute-force all $2^{k+1}$ possibilities for the $j$\ts{th} bit of the nodes in each color until one of them keeps the invariant that a cycle (resp. path) exists, continuing with $j$ until all bits are filled in. This increases the runtime by a $2^{O(k)}$ factor, and maintains the same asymptotic time (and catalytic space) complexity as \Cref{cor:paths,cor:cycles}.}

\section{Time-Space Tradeoff Lower Bounds}\label{sec:lbs}

We now discuss how the above algorithmic results relate to time-space tradeoff \emph{lower bounds}.
Although our algorithms are stated in an output model that involves XORing into the dirty tape, we show that we can instead describe them in separate-tape-output models, where existing lower bounds imply that dirty space is provably necessary for fast algorithms.

\subsection{Separating catalytic and non-catalytic time-and-space-bounded complexity}

Recall that our catalytic algorithms are stated in the \emph{dirty-XOR} output model.
We observe that they could equivalently be stated in the $2$-output-pass output model with no asymptotic loss in space or runtime.
\begin{lemma} \label{lem:normalform}
    Assume there exists a clean space $S$, catalytic space $C$, and time $T$ algorithm $A$ which computes $f=f_n \colon \set{0,1}^n \to \set{0,1}^m$ in the dirty-XOR output model.
    Then, there exists a clean space $O(S)$, dirty space $C$, time $O(T)$ algorithm which given $x$ outputs $f(x)$ on a 2-pass XOR output tape.
\end{lemma}
\begin{proof}
    First, output the first $m$ bits of the catalytic tape.
    Then, simulate $A$ to completion.
    XOR in the new first $m$ bits of the catalytic tape in a second pass.
    Finally, simulate $A$ a second time in order to reset the dirty space.
\end{proof}

We then show that, for algorithms without catalytic space, the $2$-output-pass XOR output model is equivalent to the transducer (i.e., 1-pass) output model. In fact, the proof of \Cref{lem:2passlbs} below also applies to a modified $k$-output-pass model where \emph{both} XOR-into and overwrite instructions are allowed.
\begin{lemma} \label{lem:2passlbs}
    Let $k \in \N$ be a constant. Given any space $S$, time $T$ algorithm $A$ which outputs $f=f_n \colon \set{0,1}^n \to \set{0,1}^\star$ in the $k$-output-pass output model, there exists a space $O(kS)$, time $O(T)$ algorithm which computes $f$ in a single pass.
\end{lemma}
\begin{proof}
    First, simulate $A$, ignoring its outputs.
    Every time $A$ begins a new pass, record its full configuration.
    We will thus record $k$ distinct configurations, each of which takes $O(S)$ space to write down.
    Given these configurations, we now proceed in parallel: simulate each of them until they output their first bit, then XOR all of those bits together and output them. Proceed likewise for the second bit, and so on.
\end{proof}

There are known $\Omega(n^2)$ time-space-product lower bounds on DFT and sorting, and $\Omega(n^3)$ time-space-product lower bounds on matrix multiplication, in the transducer model. By \cref{lem:2passlbs}, these lower bounds also apply to the 2-output-pass output model. So, since catalytic algorithms can achieve time--clean space product $n^{1+o(1)}$ and $n^{\omega + o(1)}$, respectively, in this model, this shows that dirty space is unconditionally helpful.

We also note that we can make a similar statement in the transducer model:

\begin{proposition}\label{prop:transducerseparation}
        There exists some relation $R \subseteq \bin^* \times \bin^*$ such that,
    \begin{itemize}
        \item there exists a 1-output-pass catalytic algorithm using clean space $O(\log(n))$, dirty space $\poly(n)$, and time $n^{1+o(1)}$, which takes in $x \in \bin^n$ and outputs some $y$ such that $(x,y) \in R$, but
        \item no (non-catalytic) 1-output-pass algorithm can output such a $y$ using $n^{o(1)}$ space and $n^{2 - \Omega(1)}$ time.
    \end{itemize}
\end{proposition}

\begin{proof}
    Consider the relation $R = \{(L, (L', L'')) \ | \ L' \oplus L'' = \mathsf{SORTED}(L)\}$. That is, given a list $L$, we want to output a pair of lists whose bitwise-XOR is equal to $L$ in sorted order.
    Observe that outputting such an $(L', L'')$ in the transducer model is equivalent to outputting the sorted list in the $2$-output-pass model.
    So, this is implied by our above results.
\end{proof}

\subsection{Separating 1-pass and 2-pass catalytic time-and-space-bounded complexity}

Given that the $k$-pass XOR and $1$-pass XOR models are equivalent for constant $k$ (up to constant factors in time and space) without dirty space, one might ask whether they are similarly equivalent for catalytic algorithms. We show that this is unconditionally false: while catalytic algorithms can quickly sort in the $2$-pass XOR model, they cannot do so in only a single pass.

\begin{theorem}
    For any catalytic transducer that takes in a length-$n$ list of $\log(n)$-bit integers and outputs the sorted list, we must have $ST \geq \Omega(n^2)$, where $S$ and $T$ are the clean space and runtime used by the algorithm, respectively.
\end{theorem}
\begin{proof}
    The proof we give directly follows Beame's time-space tradeoff lower bound in the case with no catalytic space~\cite{beame91}.
    
    Recall the notion of a \emph{catalytic branching program} operating over a list~\cite{nonuniform15}.
    A standard branching program is a directed acyclic graph, where each node is labeled with an index of the list to read, and has an outgoing edge corresponding to each possible value that could be read at each index. Some nodes are also marked with one or more \emph{output values}.
    Given an input list, one can produce an output list by starting at a designated ``input node'', querying the input list in the specified place, following the corresponding out-edge, outputting any output values specified by the new node, and repeating.    
    In a \emph{catalytic} branching program, there are instead \emph{several} nodes designated as ``start nodes'', each with a corresponding ``end node''. We require that the output is always correct regardless of which of these start nodes the program begins at, and that the program always ends at the end node corresponding to the start node where it began.
    Say that a branching program is ``layered'' if nodes can be partitioned into layers such that each edge connects nodes in adjacent layers.
    Let the \emph{length} of a branching program be the number of layers, and the \emph{amortized size} be the number of nodes divided by the number of start nodes.

    Observe that, if we have a time-$T$, clean-space $S$ catalytic transducer for a problem, we immediately get a length-$O(T)$, amortized width--$2^{O(S)}$ catalytic branching program for that problem.
    (To see this, let each possible state of the Turing machine correspond to a node of the branching program, let the edges correspond to the transitions the machine would perform, and let the start and end nodes correspond to the machine in the start and end states with the clean space empty. If $w$, $c$, and $f$ are the number of bits of workspace, catalytic space, and finite Turing machine control respectively used by the machine, then this catalytic branching program will have $2^{w + c+ f}$ nodes and $2^c$ start nodes.)
    So, it suffices to show that sorting does not have a catalytic branching program of amortized size $2^S$ and length $T$ for any $ST \leq o(n^2)$.

    In fact, we show that the \emph{unique elements} problem, of outputting (in an arbitrary order) all of the elements of a list that appear exactly once, does not have such a catalytic branching program. 
    As a time-$T$, clean-space $S$ transducer for sorting can easily be converted into a time-$O(T)$, clean-space $O(S)$ transducer for unique elements, this gives the desired result.

    Fix some catalytic branching program with $T$ layers, $x$ start vertices, and $x \cdot 2^S$ total vertices. For a node $v$ in the branching program, and an input list $L$, we will say $(v, L)$ is \emph{good} if, when starting from $v$ and taking $n/2$ branching program steps, the program outputs at least $n^2/400T$ distinct values in $[n]$, each of which appears at most once in $L$.

    \begin{claim}\label{clem:notsogood}
        For any node $v$ in any catalytic branching program, if we choose $L$ uniformly at random from $[n]^n$, we have $\Pr_L[(v, L) \text{ is good}] \leq (1 - \Omega(1))^{n^2/T}$.
    \end{claim}
    \begin{proof}
        We can imagine randomly choosing values of $L$ one-at-a-time as the branching program queries them. 
        Once $n/2$ steps have been taken, there would then remain at least $n/2$ indices of $L$ which have yet to be determined, all of which will be chosen uniformly and independently from $[n]$.
        So, for any value $y \in [n]$ that has been output after these steps, there is at least a $1 - (1 - 1/n)^{n/2} \geq \Omega(1)$ chance that $y$ appears at least twice in $L$. 
        These events are anti-correlated, so if we output more than $n^2/400T$ values, then there is at most a $(1 - \Omega(1))^{n^2/400T}$ probability that none of them appear more than once in the list.
    \end{proof}

    \begin{claim}\label{clem:gottabegood}
        Suppose a given catalytic branching program, with $x$ start vertices, correctly solves unique elements on a given list $L \in [n]^n$, and $L$ contains at least $n/100$ unique elements. Then, the branching program must contain at least $x$ vertices $v$ such that $(v, L)$ is good.
    \end{claim}
    \begin{proof}
        In order for a given computation path in the branching program to output $n/100$ correct values, it must pass through at least one vertex $v$ such that $(v, L)$ is good.
        If it didn't, then either some node along the path outputs an incorrect value, or else we can divide the computation path into $2T/n$ segments of length $n/2$, each of which outputs at most $n^2/400T$ values, meaning at most $n/200$ outputs in total.

        Now, note that, none of the computation paths from any of the start nodes can intersect each other, because otherwise one of them would fail to end up at the correct end node.
        Thus, in order for all $x$ of them to correctly solve unique elements on $L$, there must be at least $x$ good vertices.
    \end{proof}

    These two claims together imply the overall result.
    By \cref{clem:notsogood}, we have \[\mathbb{E}_{L \sim [n]^n}[\# \text{good vertices}] \leq x2^S(1- \Omega(1))^{n^2/T}=x2^{S-\Omega(n^2/T)},\] meaning by Markov's inequality that \[\Pr_{L \sim [n]^n}[\# \text{good vertices} \geq x2^{S-\Omega(n^2/T)}] < 1/2.\]
    %
    But, observe that \[\Pr_{L \sim [n]^n}[L \text{ has at least } n/100 \text{ unique elements}] > 1/2.\]
    So, there exists at least one value of $L$ such that the branching program has both fewer than $x 2^{S-\Omega(n^2/T)}$ good vertices and greater than $n/100$ unique elements.
    If we have $ST \leq o(n^2)$ (so $S \leq o(n^2/T)$), then this would mean strictly fewer than $x$ good vertices, so \cref{clem:gottabegood} would imply that the catalytic branching program is incorrect on $L$.
\end{proof}

\printbibliography

@article{clifford2007simple,
  author       = {Peter Clifford and
                  Rapha{\"{e}}l Clifford},
  title        = {Simple deterministic wildcard matching},
  journal      = {Inf. Process. Lett.},
  volume       = {101},
  number       = {2},
  pages        = {53--54},
  year         = {2007},
  url          = {https://doi.org/10.1016/j.ipl.2006.08.002},
  doi          = {10.1016/J.IPL.2006.08.002},
  bibsource    = {dblp computer science bibliography, https://dblp.org}
}

@article{datastream2026,
  author       = {Ripley Becker and
                  Sourav Chakraborty and
                  Debarshi Chanda and
                  A. Pavan and
                  N. V. Vinodchandran},
  title        = {Computing over Data Streams using Catalytic Space},
  journal      = {CoRR},
  volume       = {abs/2607.08559},
  year         = {2026},
  url          = {https://doi.org/10.48550/arXiv.2607.08559},
  doi          = {10.48550/ARXIV.2607.08559},
  eprinttype   = {arXiv},
  eprint       = {2607.08559},
  bibsource    = {dblp computer science bibliography, https://dblp.org}
}

@incollection {streaming26,
    AUTHOR = {Kaplan, Tamara and Kaplan, Nimrod and Kaplan, Haim},
     TITLE = {Streaming with catalytic memory},
 BOOKTITLE = {34th {A}nnual {E}uropean {S}ymposium on {A}lgorithms},
    SERIES = {LIPIcs. Leibniz Int. Proc. Inform.},
    VOLUME = {388},
     PAGES = {Paper No. 70, 21},
 PUBLISHER = {Schloss Dagstuhl. Leibniz-Zent. Inform., Wadern},
      YEAR = {2026},
      ISBN = {978-3-95977-445-1},
   MRCLASS = {68Q09},
  MRNUMBER = {5121559},
       DOI = {10.4230/lipics.esa.2026.70},
       URL = {https://doi.org/10.4230/lipics.esa.2026.70},
}

@article {beame91,
    AUTHOR = {Beame, Paul},
     TITLE = {A general sequential time-space tradeoff for finding unique
              elements},
   JOURNAL = {SIAM J. Comput.},
  FJOURNAL = {SIAM Journal on Computing},
    VOLUME = {20},
      YEAR = {1991},
    NUMBER = {2},
     PAGES = {270--277},
      ISSN = {0097-5397},
   MRCLASS = {68Q25 (68P10 68Q05)},
  MRNUMBER = {1087749},
MRREVIEWER = {C.\ F.\ Kelemen},
       DOI = {10.1137/0220017},
       URL = {https://doi.org/10.1137/0220017},
}

@inproceedings{alon1994color,
author = {Alon, Noga and Yuster, Raphy and Zwick, Uri},
title = {Color-coding: a new method for finding simple paths, cycles and other small subgraphs within large graphs},
year = {1994},
isbn = {0897916638},
publisher = {Association for Computing Machinery},
address = {New York, NY, USA},
url = {https://doi.org/10.1145/195058.195179},
doi = {10.1145/195058.195179},
booktitle = {Proceedings of the Twenty-Sixth Annual ACM Symposium on Theory of Computing},
pages = {326–335},
numpages = {10},
location = {Montreal, Quebec, Canada},
series = {STOC '94}
}

@article {MR819156,
    AUTHOR = {Fredman, Michael L. and Koml\'os, J\'anos and Szemer\'edi,
              Endre},
     TITLE = {Storing a sparse table with {$O(1)$} worst case access time},
   JOURNAL = {J. Assoc. Comput. Mach.},
  FJOURNAL = {Journal of the Association for Computing Machinery},
    VOLUME = {31},
      YEAR = {1984},
    NUMBER = {3},
     PAGES = {538--544},
      ISSN = {0004-5411,1557-735X},
   MRCLASS = {68P05},
  MRNUMBER = {819156},
       DOI = {10.1145/828.1884},
       URL = {https://doi.org/10.1145/828.1884},
}

@article {MR1059653,
    AUTHOR = {Schmidt, Jeanette P. and Siegel, Alan},
     TITLE = {The spatial complexity of oblivious {$k$}-probe hash
              functions},
   JOURNAL = {SIAM J. Comput.},
  FJOURNAL = {SIAM Journal on Computing},
    VOLUME = {19},
      YEAR = {1990},
    NUMBER = {5},
     PAGES = {775--786},
      ISSN = {0097-5397},
   MRCLASS = {68P05 (68P10 68Q25)},
  MRNUMBER = {1059653},
       DOI = {10.1137/0219054},
       URL = {https://doi.org/10.1137/0219054},
}

@book{Bini1994-hg,
  author       = {Dario Bini and
                  Victor Y. Pan},
  title        = {Polynomial and matrix computations, 1st Edition},
  series       = {Progress in theoretical computer science},
  volume       = {12},
  publisher    = {Birkh{\"{a}}user},
  year         = {1994},
  url          = {https://www.worldcat.org/oclc/312012822},
  isbn         = {3764337869},
  bibsource    = {dblp computer science bibliography, https://dblp.org}
}

@article{DBLP:journals/corr/Grenet25,
  author       = {Bruno Grenet},
  title        = {Fast polynomial computations with space constraints},
  journal      = {CoRR},
  volume       = {abs/2511.11267},
  year         = {2025},
  url          = {https://doi.org/10.48550/arXiv.2511.11267},
  doi          = {10.48550/ARXIV.2511.11267},
  eprinttype   = {arXiv},
  eprint       = {2511.11267},
  bibsource    = {dblp computer science bibliography, https://dblp.org}
}

@article{DBLP:journals/jsc/DumasG26,
  author       = {Jean{-}Guillaume Dumas and
                  Bruno Grenet},
  title        = {Fast in-place accumulation},
  journal      = {J. Symb. Comput.},
  volume       = {134},
  pages        = {102523},
  year         = {2026},
  url          = {https://doi.org/10.1016/j.jsc.2025.102523},
  doi          = {10.1016/J.JSC.2025.102523},
  bibsource    = {dblp computer science bibliography, https://dblp.org}
}

@inproceedings{DBLP:conf/issac/DumasG24a,
  author       = {Jean{-}Guillaume Dumas and
                  Bruno Grenet},
  editor       = {Jonathan D. Hauenstein and
                  Wen{-}shin Lee and
                  Shaoshi Chen},
  title        = {In-place fast polynomial modular remainder},
  booktitle    = {Proceedings of the 2024 International Symposium on Symbolic and Algebraic
                  Computation, {ISSAC} 2024, Raleigh, NC, USA, July 16-19, 2024},
  pages        = {26--35},
  publisher    = {{ACM}},
  year         = {2024},
  url          = {https://doi.org/10.1145/3666000.3669672},
  doi          = {10.1145/3666000.3669672},
  bibsource    = {dblp computer science bibliography, https://dblp.org}
}

@article{DBLP:journals/talg/ChanMR18,
  author       = {Timothy M. Chan and
                  J. Ian Munro and
                  Venkatesh Raman},
  title        = {Selection and Sorting in the {`Restore'} Model},
  journal      = {{ACM} Trans. Algorithms},
  volume       = {14},
  number       = {2},
  pages        = {11:1--11:18},
  year         = {2018},
  url          = {https://doi.org/10.1145/3168005},
  doi          = {10.1145/3168005},
  bibsource    = {dblp computer science bibliography, https://dblp.org}
}

@ARTICLE{in-place-FFT,
  author={Burrus, C. and Eschenbacher, P.},
  journal={IEEE Transactions on Acoustics, Speech, and Signal Processing}, 
  title={An in-place, in-order prime factor FFT algorithm}, 
  year={1981},
  volume={29},
  number={4},
  pages={806-817},
  doi={10.1109/TASSP.1981.1163645}}

@article{DBLP:journals/siamsc/Temperton91,
  author       = {Clive Temperton},
  title        = {Self-Sorting In-Place Fast Fourier Transforms},
  journal      = {{SIAM} J. Sci. Comput.},
  volume       = {12},
  number       = {4},
  pages        = {808--823},
  year         = {1991},
  url          = {https://doi.org/10.1137/0912043},
  doi          = {10.1137/0912043},
  bibsource    = {dblp computer science bibliography, https://dblp.org}
}

@article{in-place25,
  author       = {James Cook and
                  Surendra Ghentiyala and
                  Ian Mertz and
                  Edward Pyne and
                  Nathan S. Sheffield},
  title        = {The Structure of In-Place Space-Bounded Computation},
  journal      = {CoRR},
  volume       = {abs/2510.12005},
  year         = {2025},
  url          = {https://doi.org/10.48550/arXiv.2510.12005},
  doi          = {10.48550/ARXIV.2510.12005},
  eprinttype   = {arXiv},
  eprint       = {2510.12005},
  bibsource    = {dblp computer science bibliography, https://dblp.org}
}

@article{DBLP:journals/sigact/Pasanen99,
  author       = {Tomi Pasanen},
  title        = {In-place algorithms for sorting problems},
  journal      = {{SIGACT} News},
  volume       = {30},
  number       = {2},
  pages        = {61},
  year         = {1999},
  url          = {https://doi.org/10.1145/568547.568565},
  doi          = {10.1145/568547.568565},
  bibsource    = {dblp computer science bibliography, https://dblp.org}
}

@article{DBLP:journals/siamcomp/Schonhage81,
  author       = {Arnold Sch{\"{o}}nhage},
  title        = {Partial and Total Matrix Multiplication},
  journal      = {{SIAM} J. Comput.},
  volume       = {10},
  number       = {3},
  pages        = {434--455},
  year         = {1981},
  url          = {https://doi.org/10.1137/0210032},
  doi          = {10.1137/0210032},
  bibsource    = {dblp computer science bibliography, https://dblp.org}
}

@inproceedings{chmel2026frontier,
  author       = {Petr Chmel and
                  Aditi Dudeja and
                  Michal Kouck{\'{y}} and
                  Ian Mertz and
                  Ninad Rajgopal},
  editor       = {Dana Moshkovitz},
  title        = {Frontier Space-Time Algorithms Using Only Full Memory},
  booktitle    = {41st Computational Complexity Conference, {CCC} 2026, Lisbon, Portugal,
                  August 3-6, 2026},
  series       = {LIPIcs},
  volume       = {383},
  pages        = {29:1--29:23},
  publisher    = {Schloss Dagstuhl - Leibniz-Zentrum f{\"{u}}r Informatik},
  year         = {2026},
  url          = {https://doi.org/10.4230/LIPIcs.CCC.2026.29},
  doi          = {10.4230/LIPICS.CCC.2026.29},
  bibsource    = {dblp computer science bibliography, https://dblp.org}
}

@article{DBLP:journals/cc/Pyne25,
  author       = {Edward Pyne},
  title        = {Derandomizing Logspace With a Small Shared Hard Drive},
  journal      = {Comput. Complex.},
  volume       = {34},
  number       = {2},
  pages        = {13},
  year         = {2025},
  url          = {https://doi.org/10.1007/s00037-025-00275-6},
  doi          = {10.1007/S00037-025-00275-6},
  bibsource    = {dblp computer science bibliography, https://dblp.org}
}

@inproceedings{DBLP:conf/focs/AgarwalaM25,
  author       = {Aryan Agarwala and
                  Ian Mertz},
  title        = {Bipartite Matching is in Catalytic Logspace},
  booktitle    = {66th {IEEE} Annual Symposium on Foundations of Computer Science, {FOCS}
                  2025, Sydney, Australia, December 14-17, 2025},
  pages        = {360--372},
  publisher    = {{IEEE}},
  year         = {2025},
  url          = {https://doi.org/10.1109/FOCS63196.2025.00022},
  doi          = {10.1109/FOCS63196.2025.00022},
  bibsource    = {dblp computer science bibliography, https://dblp.org}
}

@inproceedings{DBLP:conf/innovations/AgarwalaAV26,
  author       = {Aryan Agarwala and
                  Yaroslav Alekseev and
                  Antoine Vinciguerra},
  editor       = {Shubhangi Saraf},
  title        = {Linear Matroid Intersection Is in Catalytic Logspace},
  booktitle    = {17th Innovations in Theoretical Computer Science Conference, {ITCS}
                  2026, Bocconi University, Milan, Italy, January 27-30, 2026},
  series       = {LIPIcs},
  pages        = {3:1--3:23},
  publisher    = {Schloss Dagstuhl - Leibniz-Zentrum f{\"{u}}r Informatik},
  year         = {2026},
  url          = {https://doi.org/10.4230/LIPIcs.ITCS.2026.3},
  doi          = {10.4230/LIPICS.ITCS.2026.3},
  bibsource    = {dblp computer science bibliography, https://dblp.org}
}

@inproceedings{DBLP:conf/issac/DumasG24,
  author       = {Jean{-}Guillaume Dumas and
                  Bruno Grenet},
  editor       = {Jonathan D. Hauenstein and
                  Wen{-}shin Lee and
                  Shaoshi Chen},
  title        = {In-place accumulation of fast multiplication formulae},
  booktitle    = {Proceedings of the 2024 International Symposium on Symbolic and Algebraic
                  Computation, {ISSAC} 2024, Raleigh, NC, USA, July 16-19, 2024},
  pages        = {16--25},
  publisher    = {{ACM}},
  year         = {2024},
  url          = {https://doi.org/10.1145/3666000.3669671},
  doi          = {10.1145/3666000.3669671},
  bibsource    = {dblp computer science bibliography, https://dblp.org}
}

@article{DBLP:journals/fttcs/ShpilkaY10,
  author       = {Amir Shpilka and
                  Amir Yehudayoff},
  title        = {Arithmetic Circuits: {A} survey of recent results and open questions},
  journal      = {Found. Trends Theor. Comput. Sci.},
  volume       = {5},
  number       = {3-4},
  pages        = {207--388},
  year         = {2010},
  url          = {https://doi.org/10.1561/0400000039},
  doi          = {10.1561/0400000039},
  bibsource    = {dblp computer science bibliography, https://dblp.org}
}

@inproceedings{DBLP:conf/innovations/CookP26,
  author       = {James Cook and
                  Edward Pyne},
  editor       = {Shubhangi Saraf},
  title        = {Efficient Catalytic Graph Algorithms},
  booktitle    = {17th Innovations in Theoretical Computer Science Conference, {ITCS}
                  2026, Bocconi University, Milan, Italy, January 27-30, 2026},
  series       = {LIPIcs},
  pages        = {43:1--43:22},
  publisher    = {Schloss Dagstuhl - Leibniz-Zentrum f{\"{u}}r Informatik},
  year         = {2026},
  url          = {https://doi.org/10.4230/LIPIcs.ITCS.2026.43},
  doi          = {10.4230/LIPICS.ITCS.2026.43},
  bibsource    = {dblp computer science bibliography, https://dblp.org}
}

@inproceedings{DBLP:conf/mfcs/AlekseevFMSV25,
  author       = {Yaroslav Alekseev and
                  Yuval Filmus and
                  Ian Mertz and
                  Alexander Smal and
                  Antoine Vinciguerra},
  editor       = {Pawel Gawrychowski and
                  Filip Mazowiecki and
                  Michal Skrzypczak},
  title        = {Catalytic Computing and Register Programs Beyond Log-Depth},
  booktitle    = {50th International Symposium on Mathematical Foundations of Computer
                  Science, {MFCS} 2025, Warsaw, Poland, August 25-29, 2025},
  series       = {LIPIcs},
  pages        = {6:1--6:18},
  publisher    = {Schloss Dagstuhl - Leibniz-Zentrum f{\"{u}}r Informatik},
  year         = {2025},
  url          = {https://doi.org/10.4230/LIPIcs.MFCS.2025.6},
  doi          = {10.4230/LIPICS.MFCS.2025.6},
  bibsource    = {dblp computer science bibliography, https://dblp.org}
}

@article{nonuniform15,
  author       = {Vincent Girard and Michal Kouck{\'{y}} and Pierre McKenzie},
  title        = {Nonuniform catalytic space and the direct sum for space},
  journal      = {Electron. Colloquium Comput. Complex.},
  volume       = {{TR15}},
  eid          = {{TR15-138}},
  year         = {2015},
  url          = {https://eccc.weizmann.ac.il/report/2015/138},
  eprinttype   = {ECCC},
  eprint       = {TR15-138},
  bibsource    = {dblp computer science bibliography, https://dblp.org}
}

@inproceedings{DBLP:conf/innovations/FolkertsmaMST25,
  author       = {Marten Folkertsma and
                  Ian Mertz and
                  Florian Speelman and
                  Quinten Tupker},
  editor       = {Raghu Meka},
  title        = {Fully Characterizing Lossy Catalytic Computation},
  booktitle    = {16th Innovations in Theoretical Computer Science Conference, {ITCS}
                  2025, Columbia University, New York, NY, USA, January 7-10, 2025},
  series       = {LIPIcs},
  pages        = {50:1--50:13},
  publisher    = {Schloss Dagstuhl - Leibniz-Zentrum f{\"{u}}r Informatik},
  year         = {2025},
  url          = {https://doi.org/10.4230/LIPIcs.ITCS.2025.50},
  doi          = {10.4230/LIPICS.ITCS.2025.50},
  bibsource    = {dblp computer science bibliography, https://dblp.org}
}

@inproceedings{DBLP:conf/focs/KouckyMPS25,
  author       = {Michal Kouck{\'{y}} and
                  Ian Mertz and
                  Edward Pyne and
                  Sasha Sami},
  title        = {Collapsing Catalytic Classes},
  booktitle    = {66th {IEEE} Annual Symposium on Foundations of Computer Science, {FOCS}
                  2025, Sydney, Australia, December 14-17, 2025},
  pages        = {455--463},
  publisher    = {{IEEE}},
  year         = {2025},
  url          = {https://doi.org/10.1109/FOCS63196.2025.00025},
  doi          = {10.1109/FOCS63196.2025.00025},
  bibsource    = {dblp computer science bibliography, https://dblp.org}
}

@inproceedings{DBLP:conf/stoc/CookLMP25,
  author       = {James Cook and
                  Jiatu Li and
                  Ian Mertz and
                  Edward Pyne},
  editor       = {Michal Kouck{\'{y}} and
                  Nikhil Bansal},
  title        = {The Structure of Catalytic Space: Capturing Randomness and Time via
                  Compression},
  booktitle    = {Proceedings of the 57th Annual {ACM} Symposium on Theory of Computing,
                  {STOC} 2025, Prague, Czechia, June 23-27, 2025},
  pages        = {554--564},
  publisher    = {{ACM}},
  year         = {2025},
  url          = {https://doi.org/10.1145/3717823.3718112},
  doi          = {10.1145/3717823.3718112},
  bibsource    = {dblp computer science bibliography, https://dblp.org}
}

@inproceedings{bckls14,
author = {Buhrman, Harry and Cleve, Richard and Kouck\'{y}, Michal and Loff, Bruno and Speelman, Florian},
title = {Computing with a full memory: catalytic space},
year = {2014},
isbn = {9781450327107},
publisher = {Association for Computing Machinery},
address = {New York, NY, USA},
url = {https://doi.org/10.1145/2591796.2591874},
doi = {10.1145/2591796.2591874},
booktitle = {Proceedings of the Forty-Sixth Annual ACM Symposium on Theory of Computing},
pages = {857–866},
numpages = {10},
location = {New York, New York},
series = {STOC '14}
}

@inproceedings{williams25,
author = {Williams, R. Ryan},
title = {Simulating Time with Square-Root Space},
year = {2025},
isbn = {9798400715105},
publisher = {Association for Computing Machinery},
address = {New York, NY, USA},
url = {https://doi.org/10.1145/3717823.3718225},
doi = {10.1145/3717823.3718225},
booktitle = {Proceedings of the 57th Annual ACM Symposium on Theory of Computing},
pages = {13–23},
numpages = {11},
location = {Prague, Czechia},
series = {STOC '25}
}

@inproceedings{cookmertz24,
author = {Cook, James and Mertz, Ian},
title = {Tree Evaluation Is in Space O(log n · log log n)},
year = {2024},
isbn = {9798400703836},
publisher = {Association for Computing Machinery},
address = {New York, NY, USA},
url = {https://doi.org/10.1145/3618260.3649664},
doi = {10.1145/3618260.3649664},
booktitle = {Proceedings of the 56th Annual ACM Symposium on Theory of Computing},
pages = {1268–1278},
numpages = {11},
location = {Vancouver, BC, Canada},
series = {STOC 2024}
}

@inproceedings{cookmertz20,
author = {Cook, James and Mertz, Ian},
title = {Catalytic approaches to the tree evaluation problem},
year = {2020},
isbn = {9781450369794},
publisher = {Association for Computing Machinery},
address = {New York, NY, USA},
url = {https://doi.org/10.1145/3357713.3384316},
doi = {10.1145/3357713.3384316},
booktitle = {Proceedings of the 52nd Annual ACM SIGACT Symposium on Theory of Computing},
pages = {752–760},
numpages = {9},
location = {Chicago, IL, USA},
series = {STOC 2020}
}

@book{arorabarak09,
author = {Arora, Sanjeev and Barak, Boaz},
title = {Computational Complexity: A Modern Approach},
year = {2009},
isbn = {0521424267},
publisher = {Cambridge University Press},
address = {USA},
edition = {1st}
}

@article{NP85,
  author    = {Jaroslav Ne{\v{s}}et{\v{r}}il and Svatopluk Poljak},
  title     = {On the complexity of the subgraph problem},
  journal   = {Commentationes Mathematicae Universitatis Carolinae},
  volume    = {26},
  number    = {2},
  pages     = {415--419},
  year      = {1985},
  url       = {http://eudml.org/doc/17394},
  mrnumber  = {MR803938},
  zbl       = {0571.05050},
}

@book{CLRS,
  author       = {Thomas H. Cormen and
                  Charles E. Leiserson and
                  Ronald L. Rivest and
                  Clifford Stein},
  title        = {Introduction to Algorithms, 3rd Edition},
  publisher    = {{MIT} Press},
  year         = {2009},
  url          = {http://mitpress.mit.edu/books/introduction-algorithms},
  isbn         = {978-0-262-03384-8},
  bibsource    = {dblp computer science bibliography, https://dblp.org}
}

@article{DBLP:journals/jacm/WilliamsW18,
  author       = {Virginia {Vassilevska Williams} and
                  R. Ryan Williams},
  title        = {Subcubic Equivalences Between Path, Matrix, and Triangle Problems},
  journal      = {J. {ACM}},
  volume       = {65},
  number       = {5},
  pages        = {27:1--27:38},
  year         = {2018},
  url          = {https://doi.org/10.1145/3186893},
  doi          = {10.1145/3186893},
  bibsource    = {dblp computer science bibliography, https://dblp.org}
}

@article{DBLP:journals/jcss/Abrahamson91,
  author       = {Karl R. Abrahamson},
  title        = {Time-Space Tradeoffs for Algebraic Problems on General Sequential
                  Machines},
  journal      = {J. Comput. Syst. Sci.},
  volume       = {43},
  number       = {2},
  pages        = {269--289},
  year         = {1991},
  url          = {https://doi.org/10.1016/0022-0000(91)90014-V},
  doi          = {10.1016/0022-0000(91)90014-V},
  bibsource    = {dblp computer science bibliography, https://dblp.org}
}

@article{Strassen73,
  author       = {Volker Strassen},
  title        = {Vermeidung von {D}ivisionen},
  journal      = {J. Reine Angew. Math.},
  volume       = {264},
  pages        = {184--202},
  year         = {1973},
  url          = {https://doi.org/10.1515/crll.1973.264.184},
  doi          = {10.1515/crll.1973.264.184}
}

@article{fortnow05,
  author       = {Lance Fortnow and
                  Richard J. Lipton and
                  Dieter van Melkebeek and
                  Anastasios Viglas},
  title        = {Time-space lower bounds for satisfiability},
  journal      = {J. {ACM}},
  volume       = {52},
  number       = {6},
  pages        = {835--865},
  year         = {2005},
  url          = {https://doi.org/10.1145/1101821.1101822},
  doi          = {10.1145/1101821.1101822},
  bibsource    = {dblp computer science bibliography, https://dblp.org}
}

@inproceedings{itairodeh77,
author = {Itai, Alon and Rodeh, Michael},
title = {Finding a minimum circuit in a graph},
year = {1977},
isbn = {9781450374095},
publisher = {Association for Computing Machinery},
address = {New York, NY, USA},
url = {https://doi.org/10.1145/800105.803390},
doi = {10.1145/800105.803390},
booktitle = {Proceedings of the Ninth Annual ACM Symposium on Theory of Computing},
pages = {1–10},
numpages = {10},
location = {Boulder, Colorado, USA},
series = {STOC '77}
}

@inproceedings{borodin1980time,
author = {Borodin, A. and Cook, S.},
title = {A time-space tradeoff for sorting on a general sequential model of computation},
year = {1980},
isbn = {0897910176},
publisher = {Association for Computing Machinery},
address = {New York, NY, USA},
url = {https://doi.org/10.1145/800141.804677},
doi = {10.1145/800141.804677},
booktitle = {Proceedings of the Twelfth Annual ACM Symposium on Theory of Computing},
pages = {294–301},
numpages = {8},
location = {Los Angeles, California, USA},
series = {STOC '80}
}

@article{Yesha1984,
  author       = {Yaacov Yesha},
  title        = {Time-Space Tradeoffs for Matrix Multiplication and the Discrete Fourier
                  Transform on any General Sequential Random-Access Computer},
  journal      = {J. Comput. Syst. Sci.},
  volume       = {29},
  number       = {2},
  pages        = {183--197},
  year         = {1984},
  url          = {https://doi.org/10.1016/0022-0000(84)90029-1},
  doi          = {10.1016/0022-0000(84)90029-1},
  bibsource    = {dblp computer science bibliography, https://dblp.org}
}

@book{Wegener1987,
  author    = {Wegener, Ingo},
  title     = {The Complexity of {B}oolean Functions},
  series    = {Wiley-Teubner Series in Computer Science},
  publisher = {John Wiley \& Sons and B. G. Teubner},
  address   = {Chichester and Stuttgart},
  year      = {1987},
  isbn      = {0-471-91555-6}
}

@article{SchonhageStrassen1971,
  author  = {Sch{\"o}nhage, Arnold and Strassen, Volker},
  title   = {Schnelle {M}ultiplikation gro{\ss}er {Z}ahlen},
  journal = {Computing},
  volume  = {7},
  number  = {3--4},
  pages   = {281--292},
  year    = {1971},
  doi     = {10.1007/BF02242355}
}

@inproceedings{Ajtai1983Sorting,
  author    = {Ajtai, Mikl{\'o}s and Koml{\'o}s, J{\'a}nos and Szemer{\'e}di, Endre},
  title     = {An $O(n \log n)$ sorting network},
  booktitle = {Proceedings of the 15th Annual ACM Symposium on Theory of Computing (STOC)},
  year      = {1983},
  pages     = {1--9},
  doi       = {10.1145/800061.808726},
  publisher = {ACM}
}

@article{Chandra1984Constant,
  author    = {Ashok K. Chandra and
               Larry J. Stockmeyer and
               Uzi Vishkin},
  title     = {Constant Depth Reducibility},
  journal   = {{SIAM} Journal on Computing},
  volume    = {13},
  number    = {2},
  pages     = {423--439},
  year      = {1984},
  url       = {https://doi.org/10.1137/0213028},
  doi       = {10.1137/0213028},
}

@inproceedings{Erickson1995,
  author       = {Jeff Erickson},
  editor       = {Kenneth L. Clarkson},
  title        = {Lower Bounds for Linear Satisfiability Problems},
  booktitle    = {Proceedings of the Sixth Annual {ACM-SIAM} Symposium on Discrete Algorithms,
                  22-24 January 1995. San Francisco, California, {USA}},
  pages        = {388--395},
  publisher    = {{ACM/SIAM}},
  year         = {1995},
  url          = {http://dl.acm.org/citation.cfm?id=313651.313772},
  bibsource    = {dblp computer science bibliography, https://dblp.org}
}

@article{AllenderKoucky2010,
  author    = {Eric Allender and Michal Kouck{\'y}},
  title     = {Amplifying Lower Bounds by Means of Self-Reducibility},
  journal   = {Journal of the ACM},
  volume    = {57},
  number    = {3},
  pages     = {14:1--14:36},
  year      = {2010},
  doi       = {10.1145/1706591.1706594}
}

@misc{cook2021borrow,
  author       = {Cook, James},
  title        = {How to borrow memory},
  year         = {2021},
  month        = jun,
  day          = {4},
  howpublished = {\url{https://www.falsifian.org/blog/2021/06/04/catalytic/}},
  note         = {Blog post.}
}

@article{CantorKaltofen91,
  author       = {David G. Cantor and
                  Erich L. Kaltofen},
  title        = {On Fast Multiplication of Polynomials over Arbitrary Algebras},
  journal      = {Acta Informatica},
  volume       = {28},
  number       = {7},
  pages        = {693--701},
  year         = {1991},
  url          = {https://doi.org/10.1007/BF01178683},
  doi          = {10.1007/BF01178683},
  bibsource    = {dblp computer science bibliography, https://dblp.org}
}

@book{BCS97,
  author    = {Peter B{\"u}rgisser and Michael Clausen and M. Amin Shokrollahi},
  title     = {Algebraic Complexity Theory},
  series    = {Grundlehren der mathematischen Wissenschaften},
  volume    = {315},
  publisher = {Springer},
  address   = {Berlin, Heidelberg},
  year      = {1997},
  doi       = {10.1007/978-3-662-03338-8}
}

\appendix
\section{Quickly converting between registers over different moduli} \label[appendix]{appendix:registers}

For the remaining part of this appendix, we will fix an arbitrary modular ring $\ringk=\Z/m\Z$ representable in $b:=\ceil{\log m}$ bits. (We further assume $O(\log b) \leq S$ for space $S$ algorithms.)
We show that we can efficiently convert between catalytic algorithms which compute the output by adding it into $N$ registers over a modular ring $\Z/m\Z$ and those which bitwise XOR the $N$-element output into dirty space (represented as $N$ concatenated $b$-bit blocks), and consequently registers over different moduli.

In the proofs below we will assume that all the $b$-bit dirty space chunks for any modular ring register represent a valid value in $\ringk$ (this is not always true when $m$ is not a power of two). We can use pre-processing and post-processing from the scheme of \cite[Appendix A]{chmel2026frontier} to assume this WLOG with only a $\poly(b)=\polylog(m)$ time multiplicative overhead.

First, we use a simple helper lemma.
\begin{lemma} \label{lem:xor-aux}
    Assume there exists an algorithm which XORs a value $v \in \set{0,1}^{M}$ onto a catalytic tape in time $T$, clean space $S$, and catalytic space $C$. Then, there exists an algorithm which adds $v_i \in \set{0,1}$ to register $R_i$ over $\ringk$ in the catalytic tape using time $T \cdot \polylog(m)$, clean space $O(S)$, and catalytic space $O(C)$.
\end{lemma}
\begin{proof}
    Allocate $M$ registers $R_i$ over $\ringk$ in the catalytic tape. First, XOR the $i$th bit of the dirty space input section into the least significant bit (LSB) of $R_i$ for all $i$. Run the catalytic algorithm XORing $v$ onto this chunk. XOR the $i$th bit into the LSB of $R_i$ again.
    This procedure keeps $R_i$ the same if $v_i=0$ and adds $\pm 1$ otherwise.
    In order to remove the negative sign, square the value added to each $R_i$ by using \cite{bckls14} transparent multiplication over $\ringk$. Run these programs in parallel so only $O(1)$ calls to the above $(0,\pm 1)$ procedure are used.
    \qedhere 
\end{proof}

We can now show register modular ring output implies XOR output.
\begin{lemma}[MOD $m$ to XOR] \label{lem:mod-to-xor}
Assume there exists an algorithm $\mathcal{A}$ adding vector $v \in \ringk^N$ onto catalytic tape registers in time $T$, clean space $S$, and catalytic space $C$. Then, there exists an algorithm which bitwise XORs $v$ onto a block of catalytic tape
in time $T \cdot \polylog(m)$, clean space $O(S)$, and catalytic space $O(C)$.
\end{lemma}
\begin{proof}
Suppose the $Nb$-bit output dirty space block algorithm $\mathcal{A}$ uses is initialized to $\tau \in \set{0,1}^{Nb}$. Let $\tau_i \in \set{0,1}^b$ represent the initial $b$-bit value at register $R_i$. Observe that by running $\mathcal{A}$ appropriately, we can toggle between having $a:=(\tau_i)_{i \in [N]}$ or $c:=(\tau_i+v_i)_{i \in [N]}$ in the catalytic tape.\footnote{To revert from $(\tau_i+v_i)_{i \in [N]}$ to $(\tau_i)_{i \in [N]}$ simply negate all the $N$ registers, run $\mathcal{A}$, and negate all the registers again. This is a standard trick to get the inverse (i.e. subtract).} In particular, we have read-access (and therefore XOR-access) to these values.

Allocate $2Nb$ registers over $\Z/2^k\Z$ (for a $k$ discussed shortly). Add the $j$th bit of $a$ and $c$ onto $R_{2j}$ and $R_{2j+1}$ respectively for all $j$ using \Cref{lem:xor-aux}, while only running $\mathcal{A}$ a constant number of times. (We can XOR the value of $a$ all at once onto any place in the catalytic tape. We can similarly do this for $c$ by toggling.)
There exists an $NC^1$ circuit which given two values $a_i, c_i \in \ringk$ finds $c_i-a_i$ over $\ringk$.
Let $k$ be the value needed for the power-of-two registers used in \Cref{thm:circeval} for our circuit. Using \Cref{thm:circeval}, evaluate $N$ instances of this circuit in parallel in time $O(T + N \cdot \poly(b))$ and $O(\log b) \leq O(S)$ additional space. As the output of the circuit evaluation will be in a bitwise XOR form, we can XOR the value $(c_i-a_i)_{i \in [N]}=(v_i)_{i \in [N]}$ onto a desired block.
\qedhere
\end{proof}

\paragraph{XOR-output to modular output} \label{lem:xor-to-mod} We observe it is possible to go from XOR output to modular ring output with only a $\poly(b)$ overhead as well. This follows from first applying \Cref{lem:xor-aux} to add an XOR-access bit into a corresponding register in $\ringk$ ($Nb$ registers total); then, for each output ring register ($N$ total), accumulate the sum of the $k$th least significant bit register multiplied by $2^k$ over all $k$. (Do this in parallel over all output registers to maintain only a $\polylog(m)$ overhead.)

\paragraph{Modulo $m$ output to modulo $a$ output} We additionally observe that we can convert from an algorithm which adds values to registers mod $m$ to an algorithm which subsequently takes the remainder mod some other value.

\begin{lemma}[MOD $m$ to MOD $a$] \label{lem:mod-to-mod}
    Let $m,a$ be positive integers where $1<a<m$, and $v \in \Z^N$ be a vector where each entry has magnitude at most $a$ for all $i \in [N]$.
    Assume there exists an algorithm adding vector $v$ onto catalytic tape registers over $\Z/m\Z$ in time $T$, clean space $S$, and catalytic space $C$. There exists an algorithm running in time $T\polylog(m)$, clean space $O(S)$, and catalytic space $O(C + N\polylog(m))$ which instead adds $v$ over $\Z/a\Z$.
\end{lemma}
\begin{proof}
    First, convert the added output over $\Z/m\Z$ registers to XOR form by \Cref{lem:mod-to-xor}. Then, convert to $\Z/a\Z$ register form by the previous XOR to modular register output remark.
\end{proof}

\end{document}